\documentclass[fleqn,usenatbib]{mnras}

\usepackage{newtxtext,newtxmath}
\usepackage[T1]{fontenc}
\usepackage{ae,aecompl}

\usepackage{graphicx}	% Including figure files
\usepackage{amsmath}	% Advanced maths commands
\usepackage{pdflscape}
\usepackage{booktabs}
\usepackage{multirow}

\usepackage{xcolor}
\usepackage[flushleft]{threeparttable}

\usepackage{tabularx}

\def\NII{[N\,{\sc ii}]}
\def\ArIV{[Ar\,{\sc iv}]}
\def\OIII{[O\,{\sc iii}]}

\def\FeII{[Fe\,{\sc ii}]}
\def\FeVII{[Fe\,{\sc vii}]}

\def\FeX{[Fe\,{\sc x}]}
\def\FeXI{[Fe\,{\sc xi}]}

\def\SVII{[S\,{\sc vii}]}
\def\SiVI{[Si\,{\sc vi}]}

\def\SIX{[S\,{\sc ix}]}

\def\SVIII{[S\,{\sc viii}]}
\def\SXII{[S\,{\sc xii}]}
\def\SII{[S\,{\sc ii}]}
\def\SIII{[S\,{\sc iii}]}
\def\SiX{[Si\,{\sc x}]}

\title[A study of CLiF AGNs]{Coronal Line Forest AGN I: physical properties of the emission-line regions}

\author[F. C. Cerqueira-Campos et al.]{
F. C. Cerqueira-Campos,$^{1}$\thanks{E-mail: fernando.campos@inpe.br}
A. Rodr\'iguez-Ardila,$^{1,2}$\thanks{Visiting Astronomer at the Infrared Telescope Facility, which is operated by the University of Hawaii under contract NNH14CK55B with the National Aeronautics and Space Administration.}
R. Riffel,$^3$
M. Marinello,$^{2}$
\newauthor{A. Prieto$^{4}$ and L. G. Dahmer-Hahn$^{2}$}
\\
$^{1}$Divis\~ao de Astrof\'isica, Instituto Nacional de Pesquisas Espaciais, Avenida dos Astronautas 1758, S\~ao Jos\'e dos Campos, 12227-010, SP, Brazil\\
$^{2}$Laborat\'orio Nacional de Astrof\'isica - Rua dos Estados Unidos 154, Bairro das Na\c c\~oes. CEP 37504-364, Itajub\'a, MG, Brazil\\
$^{3}$ Departamento de Astronomia, Instituto de F\'{\i}sica, Universidade Federal do Rio Grande do Sul, Av. Bento Gon\c calves, 9500. Porto Alegre, RS, Brazil.\\
$^{4}$Instituto de Astrof\'{\i}sica de Canarias, C/ Via L\'actea, S/N, San Cristobal de La Laguna, Tenerife, Spain
}

\date{Accepted XXX. Received YYY; in original form ZZZ}

\pubyear{2019}

\begin{document}
\label{firstpage}
\pagerange{\pageref{firstpage}--\pageref{lastpage}}
\maketitle

\begin{abstract}
Coronal-Line Forest Active Galactic Nuclei (CLiF AGN)  are characterized by strong high-ionization lines, which contrast to what is found in most AGNs. Here, we carry out a multiwavelength analysis aimed at understanding the physical processes in the Narrow Line Region (NLR) of these objects and unveiling if they are indeed a special class of AGN. By comparing coronal emission-line ratios we conclude that there are no differences between CLiF and non-CLiF AGNs. We derive physical conditions of the narrow line region (NLR) gas and found electron densities in the range $3.6\times$10$^{2}$ - $1.7\times$10$^{4}$ cm$^{-3}$ and temperatures of $3.7\times$10$^{3}$ - $6.3\times$10$^{4}$ K, suggesting that the ionization mechanism is associated primarily with photoionization by the AGN. We suggest a NLR dominated by matter-bounded clouds to explain the high-ionization line spectrum observed. The mass of the central black hole, derived from the stellar velocity dispersion show that most of the objects have values in the interval 10$^{7-8}$~M$\odot$.  Our results imply that CLiF AGN is not a separate category of AGNs. In all optical/near-infrared emission-line properties analyzed, they represent an extension to the low/high ends of the distribution within the AGN class. 

%are indeed special due to the enhancement of the forbidden high-ionization line spectrum, rarely observed in AGN samples. 
\end{abstract}


%%%%%%%%%%%%%%%%%%%%%%%%%%%%%%%%%%%%%%%%%%%%%%%%%%

%%%%%%%%%%%%%%%%% BODY OF PAPER %%%%%%%%%%%%%%%%%%

\section{Introduction}\label{ch:cap1}

Coronal lines (CLs), or forbidden high ionization emission lines (FHILs), originate from forbidden fine structure transitions excited by collisions in highly ionized species (ionization potential, IP $\geq $ 100 eV). Because of their very high ionization potential, they are considered to be a reliable signature of the presence of an AGN in galaxies (\citealt{penston/1984}; \citealt{prieto/2000}; \citealt{prieto/2002}; \citealt{reunanen/2003};
\citealt{satyapal/2008}; \citealt{goulding/2009}). However, they are also produced in supernova remnants (SNRs; \citealt{oliva/1999}; \citealt{smith/2009}), planetary nebulae (\citealt{pottasch/2009}), and
Wolf–Rayet stars (\citealt{schaerer/1999}). Typical CL luminosities in these latter objects are low,  $\sim$10$^{31-33}$~erg\,s$^{-1}$ (\citealt{dopita/2018}). Thus, several thousands of them might be necessary to produce detectable CL emission if an AGN origin is ruled out, and in that case  the sources need to be forming stars very actively. Moreover, in SNRs CLs strongly vary with time. \cite{komossa/2008}, \cite{komossa/2009} and \cite {wang/2011} showed that in these objects, they fade on timescales of a few years.

The presence of coronal lines in active galactic nuclei (AGN) spectra is attributed to the existence of highly energetic processes associated with the nuclear activity. Some of these lines are conspicuous and their emission region can be extended to scales of up to a few hundred parsecs (\citealt{prieto/2005}; \citealt{rodriguez/2006}; \citealt{mazzalay/2010}; \citealt{muller/2011}). Usually, [\ion{Ne}{v}]~$\lambda$3435~\AA\ (IP = 100~eV) and [\ion{Fe}{vii}]~$\lambda$6087~\AA\ (IP = 100~eV) are the most prominent CLs in the optical region \citep{murayama/1998} while [\ion{Si}{vi}]~1.963~$\mu$m (IP = 166~eV) is the strongest CL in the near-infrared (NIR). However, CL up to $\chi$=505~eV ([\ion{S}{xii}]~$\lambda$7609~\AA) have been reported in AGN \citep{mazzalay/2010}. The observed luminosities of such lines cover about three orders of magnitude, from 10$^{38}$ up to 10$^{41}$~erg\,s$^{-1}$ \citep{rodriguez/2011}.

Although CLs are frequent in AGN spectra, not all AGN display them.  \citet{riffel/2006},  using a sample of 47 AGN in the near-infrared (NIR) found that in 67\% of the objects, one CL is identified. They showed that the lack of CLs in the remaining 33\% of objects is genuine and not driven by sensitivity detection issues. In their analysis, the faintest [\ion{Si}{vi}] coronal luminosity was $\sim10^{38}$~erg\,s$^{-1}$, measured in NGC\,4051 at a redshift $z=$0.00234. More distant sources, such as Ark\,564 ($z=$0.0247) or PG\,1448+273 ($z=$0.065) display a [\ion{Si}{vi}] luminosity nearly 2 dex higher. All objects were observed with similar signal to noise ratio (S/N). Similar results are also found by \citet{lamperti/2017}. In the optical region, CLs of iron (i.e. [\ion{Fe}{vii}] and [\ion{Fe}{x}]) are usually observed, but no quantitative studies of its frequency have been made yet. Despite its importance, these lines tend to be weak. [\ion{Fe}{vii}]~6087~\AA, the brightest CL in the 3500 - 7500~\AA\ region, is usually 1-10\% the strength of [\ion{O}{iii}]~5007~\AA\ \citep{murayama/1998}.

Recently, a new class of AGN, named AGNs with coronal line forest (CLiF AGNs), were introduced in the literature (\citealt{rose/2015a}; \citealt{rose/2015b}; \citealt{glidden/2016}). Among several interesting properties,  CLiF AGN are characterized by: ($i$) the emission line flux ratio [\ion{Fe}{vii}]~$\lambda$6087/H$\beta > 0.25$; ($ii$) [\ion{Fe}{x}]~$\lambda$6374/H$\beta >$ 0.2; ($iii$) [\ion{Ne}{v}]~$\lambda$3426/H$\beta >$ 1; ($iv$) The CLs are not blueshifted with respect to the low-ionization lines ($\Delta$v $\lesssim$ 100 km\,s$^{-1}$);
($v$) The velocity widths of the CLs are narrow and similar to that of low-ionization lines (FWHM $<$ 300 km\,s$^{-1}$) using single Gaussian fits. Optical luminosities of these CLs are found to be between 10$^{40}$ to 10$^{41}$ erg\,s$^{-1}$ (see Sect.~4.4). These values are well within the interval of CL luminosities reported by \citet{gelbord/2009} in a sample 63 AGN.

Another peculiar characteristic identified by \defcitealias{rose/2015a}{RET15}\citet[][hereafter RET15]{rose/2015a} in the CLiF AGNs is the high value of the H$\alpha$/H$\beta$ ratio, which in all objects studied was between 3.8 and 6.6 (taking into account the limits of the error bars). That is, higher than the Balmer decrement assuming case B for low density gas. They interpreted this result as due to high density gas in the narrow line region instead of being attributed to dust extinction because the values found for H$\gamma$/H$\beta$ are consistent with the intrinsic case B value of 0.47 \citep{osterbrock/2006}.
 
The results obtained by \citetalias{rose/2015a} suggest that most of the H$\alpha$ emission is produced in the same region of the coronal line forest, enhancing the flux ratio H$\alpha$/H$\beta$. The gas density they obtained by modelling the NLR is in the range $10^{6}-10^{7.5} $cm$^{-3}$, much higher than that expected for typical NLR conditions (< $10^{4}$ cm$^{-3}$  \citealt{osterbrock/2006}).

\citetalias{rose/2015a} classified only 7 CLiF AGNs from an initial sample of $\sim$5000 AGN spectra from the Sloan Digital Sky Survery (SDSS).

Here, we carry out the first near-infrared study of CLiF AGN in the literature along with a re-analysis of the optical spectra of these objects. We detail in Sect. \ref{dados} the sample and observations. Our additional data widen the number and species of coronal lines suitable to derive the physical conditions of the coronal and narrow line region in these interesting sources (Sects \ref{continuum} \& \ref{lines}). Stellar and gas kinematics properties are derived in Sect. \ref{kinematics}. We also derive additional properties of CLiF AGN such as the mass of the SMBH in order to construct a more complete picture of these sources. Main conclusions are drawn in Sect. \ref{conclusions}.

\begin{table*}
\centering

\caption{Sample of CLiF AGN as defined by \protect\cite{rose/2015a} and employed in this work. Column~9 lists the classification into AGN of Type~I or~II according to \protect\cite{rose/2015b} while the last column list the classification based on results obtained from this work (see Sect. 4.1).}
\begin{tabular}{lccccccccc}
\hline
\multicolumn{1}{c}{Galaxy} & Redshift & Telescope/Instrument & Date       & E(B-V)$_{\rm G}$ & Airmass & Exp. Time     & PA$^{\circ}$ & Classification & Classification \\
                           & (z)      &                      & dd.mm.yyyy   & mag              &         & (s)           & (E of N)     &          \protect\cite{rose/2015b}   & this work      \\ \hline
ESO138-G001                & 0.0091   & SOAR/Goodman         & 11.03.2017 & 0.176            & 1.13    & 3$\times$900  & 279          & Type-2         & Type-1         \\
                           &          & Blanco/ARCoIRIS      & 07.04.2017 &                  & 1.16    & 16$\times$180 & 0.0          &                &                \\
SDSS J164+43   & 0.2210   & Gemini/GNIRS         & 15.02.2017 & 0.01             & 1.24    & 16$\times$180 & 90           & Type-2         & Type-1         \\
III Zw 77                  & 0.0342   & IRTF/SpeX            & 12.04.2015 & 0.09             & 1.08    & 18$\times$180 & 338          & Type-1         & Type-1         \\
MRK 1388                   & 0.0213   & IRTF/SpeX            & 04.05.2015 & 0.03             & 1.00    & 8$\times$200  & 300          & Type-2         & Type-1         \\
SDSS J124+44   & 0.0420   & ...         & ... & 0.02             & ...    & ... & ...           & Type-2         & Type-2         \\
2MASX J113+16    & 0.1740   & Gemini/GNIRS         & 20.02.2017 & 0.03             & 1.25    & 12$\times$160 & 90           & Type-2         & Type-1         \\
NGC 424                    & 0.0118   & SOAR/Goodman         & 07.08.2014 & 0.01             & 1.02    & 4$\times$900  & 60           & Type-2         & Type-1         \\
                           &          & Blanco/ARCoIRIS      & 19.09.2016 &                  & 1.02    & 12$\times$180 & 270          &                &                \\ \hline
\end{tabular}

\label{tab:dados}%
\end{table*}

\section{Sample, observations and data reduction}\label{dados}

The sample chosen for this analysis consists of the seven CLiF AGN already identified by \citetalias{rose/2015a}, observed using optical (3400 $\leq$ $\lambda$ $\leq$ 7500  \text{\AA}) along with NIR (7800 $\leq$ $\lambda$ $\leq$ 25000  \text{\AA}) spectroscopy. Table~\ref{tab:dados} list basic properties of the objects as well as the log of observations. Details about telescope setup and data reduction according to the wavelength interval are given below. Note that we did not collect optical spectra for 2MASX\,J113111.05+162739 nor NIR spectra for SDSS\,J124134.25+442639.2.

\subsection{Optical Spectroscopy}

Optical spectra for SDSS\,J124134.25+442639.2, MRK\,1388, III\,Zw\,77 and SDSS\,J164126.91+432121.5 were obtained from the Sloan Digital Sky Survey  Data Release 7 (SDSS DR7) database \citep{abazajian/2009}. These spectra have already been published by \citetalias{rose/2015a}. Below, we detailed the observations and data reduction procedure for ESO~138-G001 and NGC\,424.

\subsubsection{Goodman/SOAR Observations}
Optical spectra of ESO~138-G001 and NGC\,424 were obtained with the 4.1 m Southern Observatory for Astrophysical Research (SOAR) Telescope at Cerro Pachon, Chile. The observations were carried out using the Goodman Spectrograph \citep{clemens/2004}, equipped with a 400~l/mm grating and a 0.8 arcsec slit width, giving a resolution R$\sim$1500. 
In addition to the science frames, standard stars \citep{baldwin/1984} were observed for flux calibration. HgAr arc lamps were taken after the science frames for wavelength calibration. Daytime calibrations include bias and flat field images.
 
The data were reduced using standard IRAF tasks. It includes subtraction of the bias level and division of the science and flux standard star frames by a  normalized master flat-field image. Thereafter, the spectra were wavelength calibrated by applying 	the dispersion solution obtained from the arc lamp frames. 
Finally, the spectra of standard stars were extracted and combined to derived the sensitivity function, later applied to the science 1D spectra.  The final products are wavelength and flux calibrated optical spectra.

In all cases above,  the final spectra were corrected for Galactic extinction using the extinction maps of \citet{schlafly/2011} (see column~6 of Table~\ref{tab:dados}) and the \citet{cardelli/1989} law. The final reduced spectra are shown in Fig. \ref{fig:SOAR}.

\subsection{NIR Spectroscopy}

NIR of the sample listed in Table~\ref{tab:dados} were obtained at three different observatories: Cerro Tololo Inter-American Observatory (CTIO), Gemini North, and the NASA Infrared Telescope Facility (IRTF). In all cases, a cross-dispersed spectrograph, providing simultaneous coverage of the 0.8 -- 2.4~$\mu$m region was employed. Below we provide basic description of the observations and data reduction procedure for each set of data.

\subsubsection{GNIRS/Gemini spectroscopy}

Near-infrared spectra of  SDSS~J124134.25+442639.2 and SDSS~J164126.91+432121.5 were obtained in queue mode with the 8.1~m Gemini North telescope atop Mauna Kea (PROGRAM ID=GN2017A-Q40). 
The Gemini Near-IR spectrograph \citep[GNIRS,][]{elias/2006} in the cross-dispersed mode was employed. It allows simultaneous z+J, H and K band observations, covering the spectral range 0.8\,$-$\,2.5$\mu$m in a single exposure.
GNIRS science detector consist of an ALADDIN 1k $\times$ 1k InSb array. The instrument setup includes a 32~l/mm grating and a 0.8$\times7$ arcsec slit, giving a spectral resolution of R$\sim$1300 (or 320~km\,s$^{-1}$ FWHM). Individual exposures were taken, nodding the source in a ABBA pattern along the slit.
Right after the observation of the science frames, an A0V star was observed at a similar airmass, with the purpose of flux calibration and telluric correction.
 
The NIR data were reduced using the XDGNIRS pipeline (v2.0)	\footnote[1]{Based on the Gemini IRAF packages}, which delivers a full reduced, wavelength and flux calibrated, 1D spectrum with all orders combined \citep{mason/2015}. 
Briefly, the pipeline cleans the 2D images from radiative events and prepares a master flat constructed from quartz IR lamps to remove pixel to pixel variation. 
Thereafter, The s-distortion solution is obtained from daytime pinholes flats and applied to the science and telluric images to rectify them. 
Argon lamp images are then used to find the wavelength dispersion solution, followed by the extraction of 1D spectra from the combined individual exposures. 
The telluric features from the science spectrum are removed using the spectrum of a A0V star. 
Finally, the flux calibration is achieved assuming a black body shape for the  standard star \citep{pecaut/2013} scaled to its K-band magnitude \citep{skrutskie/2006}. 
The different orders are combined in to a single 1D spectrum and  corrected for Galactic extinction using the \citet{cardelli/1989} law and the extinction maps of \citet{schlafly/2011}. The spectra are shown in Fig. \ref{fig:gemini}.

\subsubsection{IRTF/SpeX data}

NIR spectra of III\,Zw~77 and Mrk\,1388 were obtained at the NASA 3~m Infrared
Telescope Facility (IRTF) using the SpeX spectrograph \citep{rayner/2003} in the short
cross-dispersed mode (SXD, 0.7-2.4~$\mu$m). The detector consists of a Teledyne 2048 $\times$ 2048 Hawaii-2RG array with a spatial
scale of 0.10 arcsec/ pixel. A 0.8 arcsec x 15 arcsec slit, providing a spectral resolution, on average, of 320~km\,s$^{-1}$ was employed.

Observations were done nodding in two positions along the slit. Right before or after the science target, a telluric star, close in airmass to the former, was observed to remove telluric features and to perform the flux calibration. CuHgAr frames were also observed at the same position as the galaxies for wavelength calibration. The spectral reduction, extraction and wavelength calibration procedures were performed using Spextool V4.1, an IDL-based software developed and provided by the SpeX team for the IRTF community \citep{cushing/2004}. Telluric features removal and flux calibration were done using XTELLCOR \citep{vacca/2003}, another software available by the SpeX team. The different orders were merged into a single 1D spectrum from 0.7 to 2.4~$\mu$m using the XMERGEORDERS routine. After this procedure, the IDL routine Xlightloss, also written by the SpeX team, was employed. It corrects an input spectrum for slit losses relative to the standard star used for flux calibration. This program is useful if either the object or the standard were not observed with the slit at the parallactic angle. Differential refraction is, indeed, the main source of uncertainty in flux calibration and it was minimised following the above
procedure. Finally, the spectra were corrected for Galactic extinction using the \citet{cardelli/1989} law and the extinction maps of \citet{schlafly/2011}. The spectra are shown in Fig. \ref{fig:irtf}.

\subsubsection{Blanco/ARCoIRIS data}

NIR spectra of ESO~138-G001 and NGC\,424 were obtained using the ARCoIRIS spectrograph attached to the 4.1~m Blanco Telescope \citep{schlawin/2014}. The science detector employed is a 2048 $\times$ 2048 Hawaii-2RG HgCdTe array with a sampling of 0.41 arcsec/pixel. The slit assembly is 1.1 arcsec wide and 28 arcsec long. The delivered spectral resolution R is $\sim$3500 across the six science orders. Similar to the IRTF/SpeX data, telluric standards were observed close to the science targets to warrant good telluric band cancellations and flux calibration. 

Data reduction was carried out using Spextool V4.1 with some modifications specifically designed for the data format and characteristics of ARCoIRIS. The procedure follows the same steps and recipes as for SpeX. The final data product consist of a 1D spectrum, wavelength and flux calibrated with all individual orders merged forming a continuous 0.9 -- 2.4~$\mu$m spectrum. We then corrected these data for Galactic extinction using the \citet{cardelli/1989} law and the extinction maps of \citet{schlafly/2011}.

In order to verify the quality of the flux calibrations in the NIR spectra, we compared our data with the 2MASS photometric points. The only data that required a calibration adjustment were those taken with ARCoIRIS (ESO~138-G001 and NGC\,424). 
In order to re-scale the flux level to that of 2MASS, a third-degree polynomial function  was applied using the Python LMFIT routine (\citealt{newville/2015}). The spectra are shown in Fig. \ref{fig:blanco}.

\begin{figure*}

    \begin{center}
    \includegraphics[width=\textwidth,height=21cm]{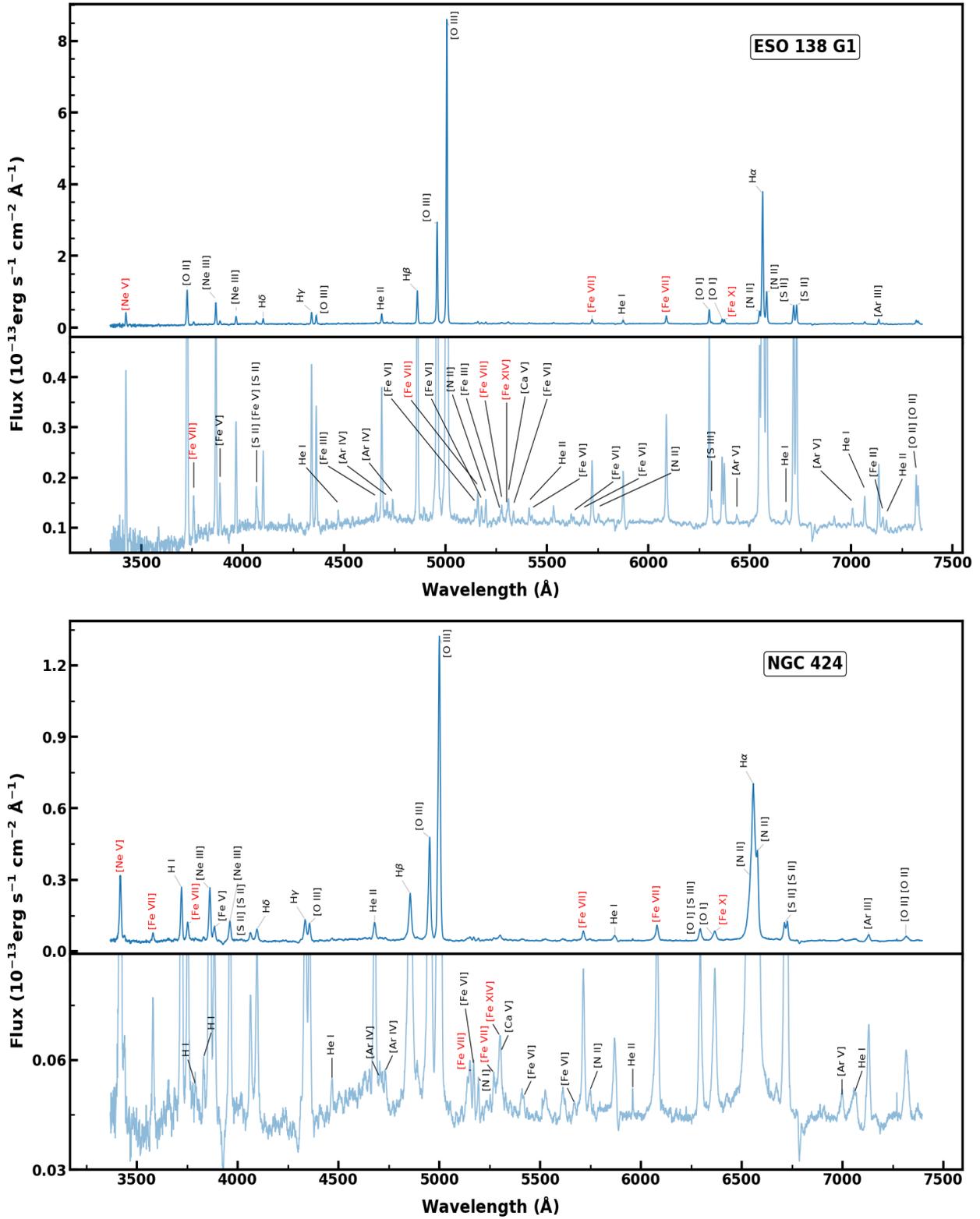}
    \caption{Optical spectra obtained by Goodman/SOAR. In the bottom panel for both spectra a zoom around the continuum level is displayed to show the faintest lines identified. The ions belonging to coronal line transitions are represented in red.}\label{fig:SOAR}
    \end{center}
	
	\centering
\end{figure*} 

\begin{figure*}

    \begin{center}
    \includegraphics[width=\textwidth,height=21cm]{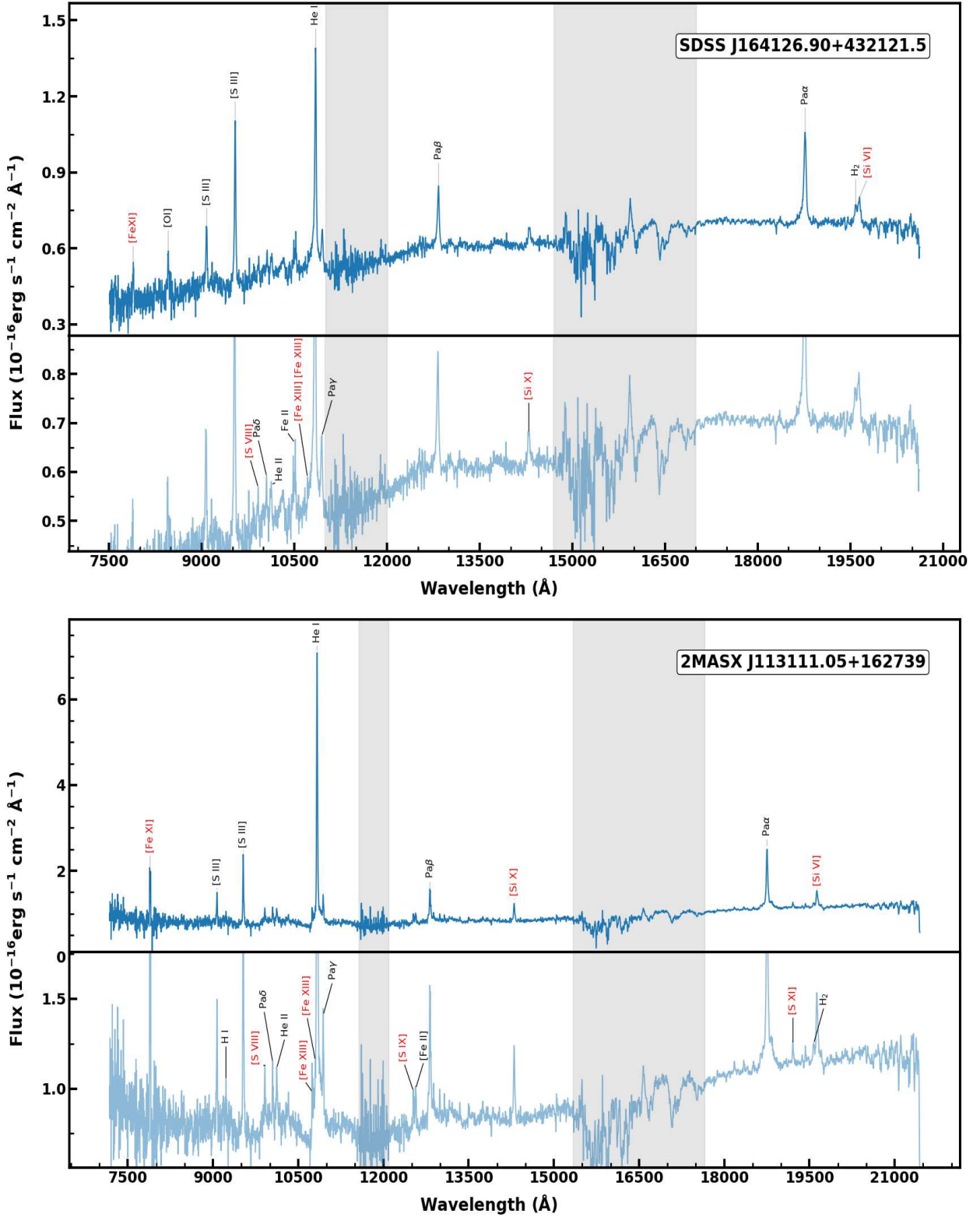}
    \caption{NIR spectra obtained by GNIRS/Gemini. The shaded area represents the region of low atmospheric transmission. Lines located in these regions were not employed in the analysis. In the bottom panel, for both spectra, a zoom is displayed to identify the faintest lines. Coronal lines are identified with red labels.}\label{fig:gemini}
    \end{center}
	
	\centering
\end{figure*} 
 
\begin{figure*}

    \begin{center}
    \includegraphics[width=\textwidth,height=21cm]{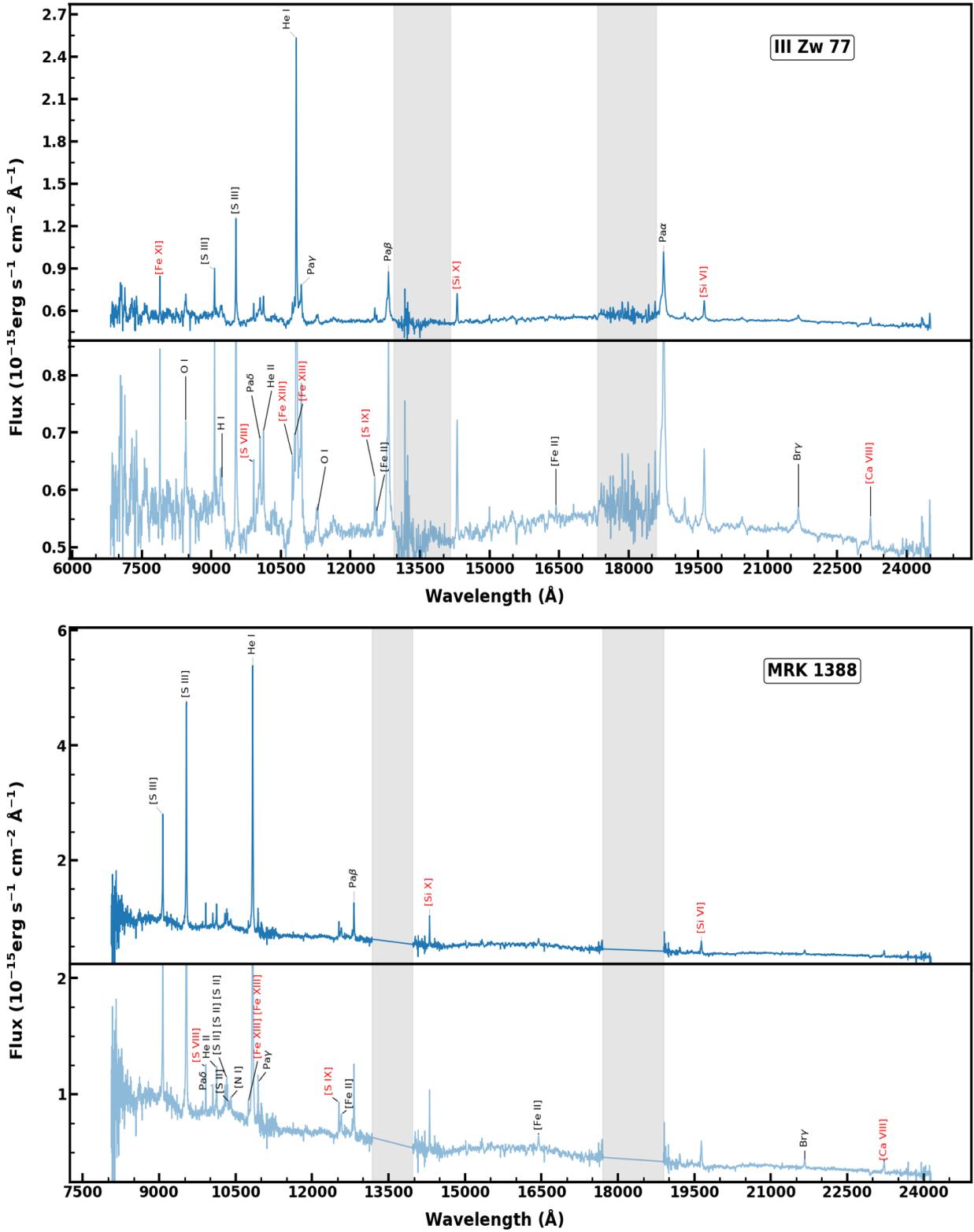}
    \caption{Same as Fig. \ref{fig:gemini} for IRTF/Spex spectra}\label{fig:irtf}
    \end{center}
	
	\centering
\end{figure*} 

\begin{figure*}

    \begin{center}
    \includegraphics[width=\textwidth,height=21cm]{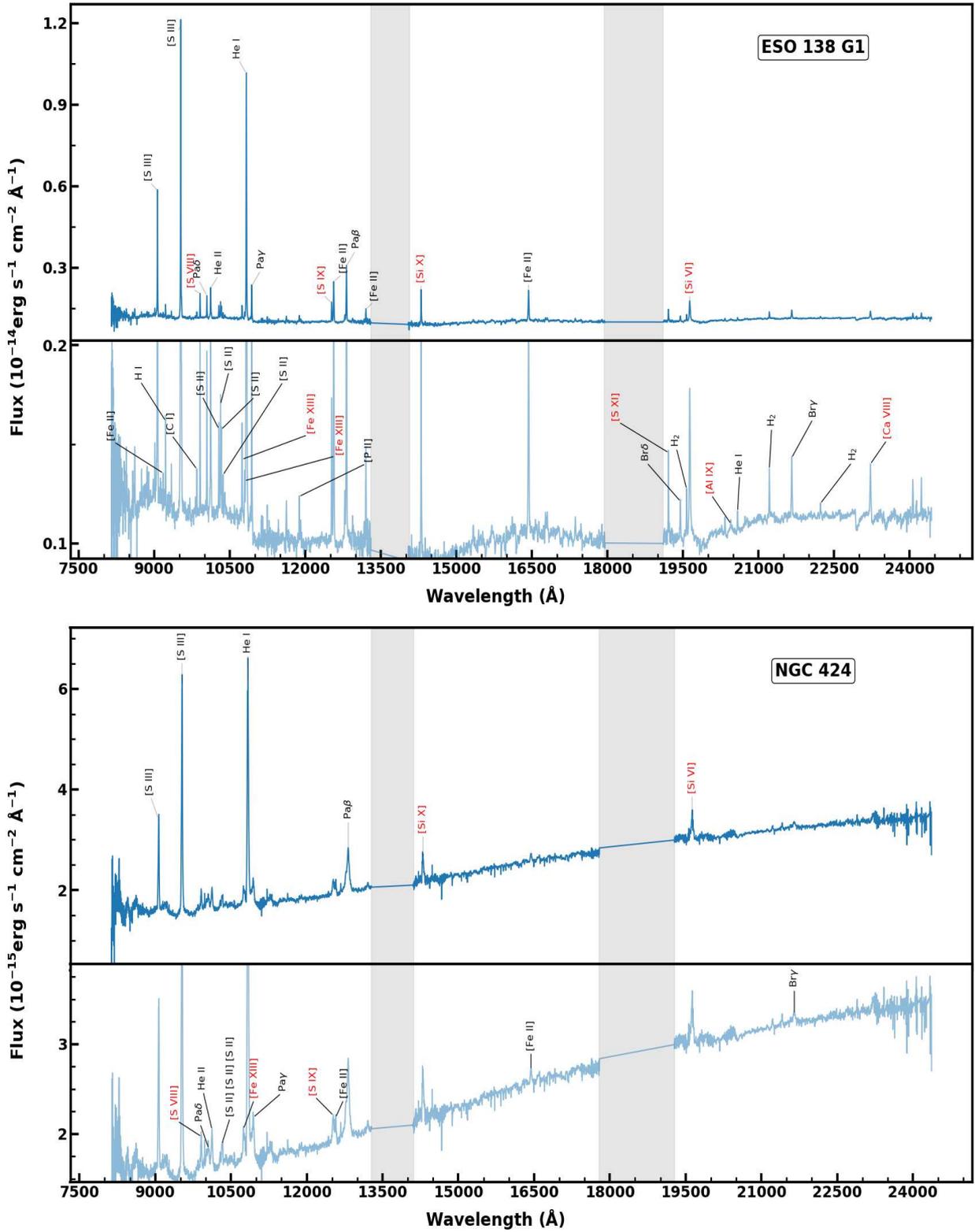}
    \caption{Same as Fig. \ref{fig:gemini} for Blanco/ARCoIRIS spectra}\label{fig:blanco}
    \end{center}
	
	\centering
\end{figure*} 

\begin{figure*}

    \begin{center}
    \includegraphics[width=\textwidth,height=13cm]{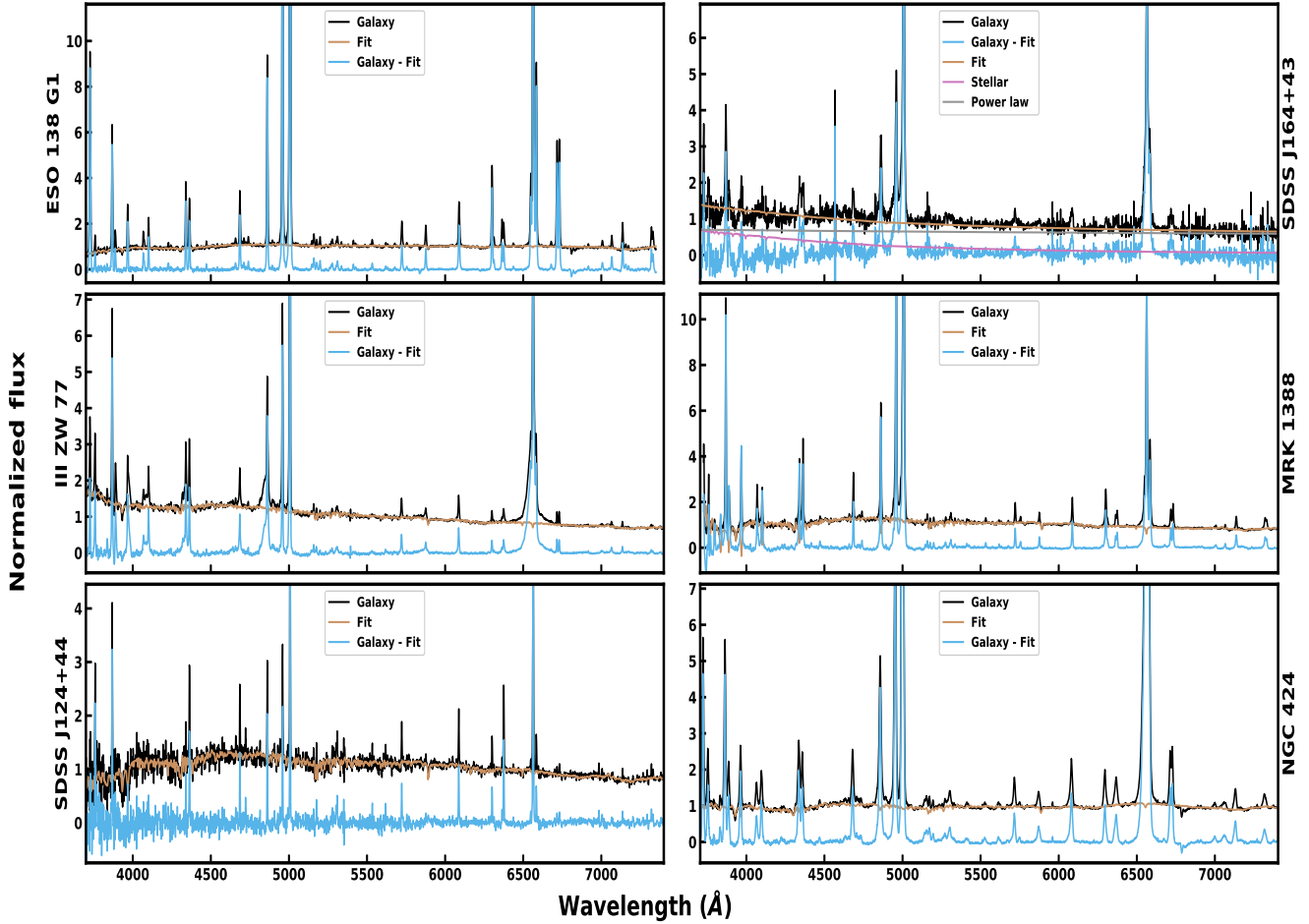}
    \caption{Modelling of the stellar contribution for CLiF AGN in the optical region. In all panels, the observed spectrum is in black, the stellar continuum is in red and the nebular emission obtained after subtraction of the stellar continuum is in blue.}\label{fig: ppxfFig_opt}
    \end{center}
	
	\centering
\end{figure*} 
 
\begin{figure*}

    \begin{center}
    \includegraphics[width=\textwidth,height=13cm]{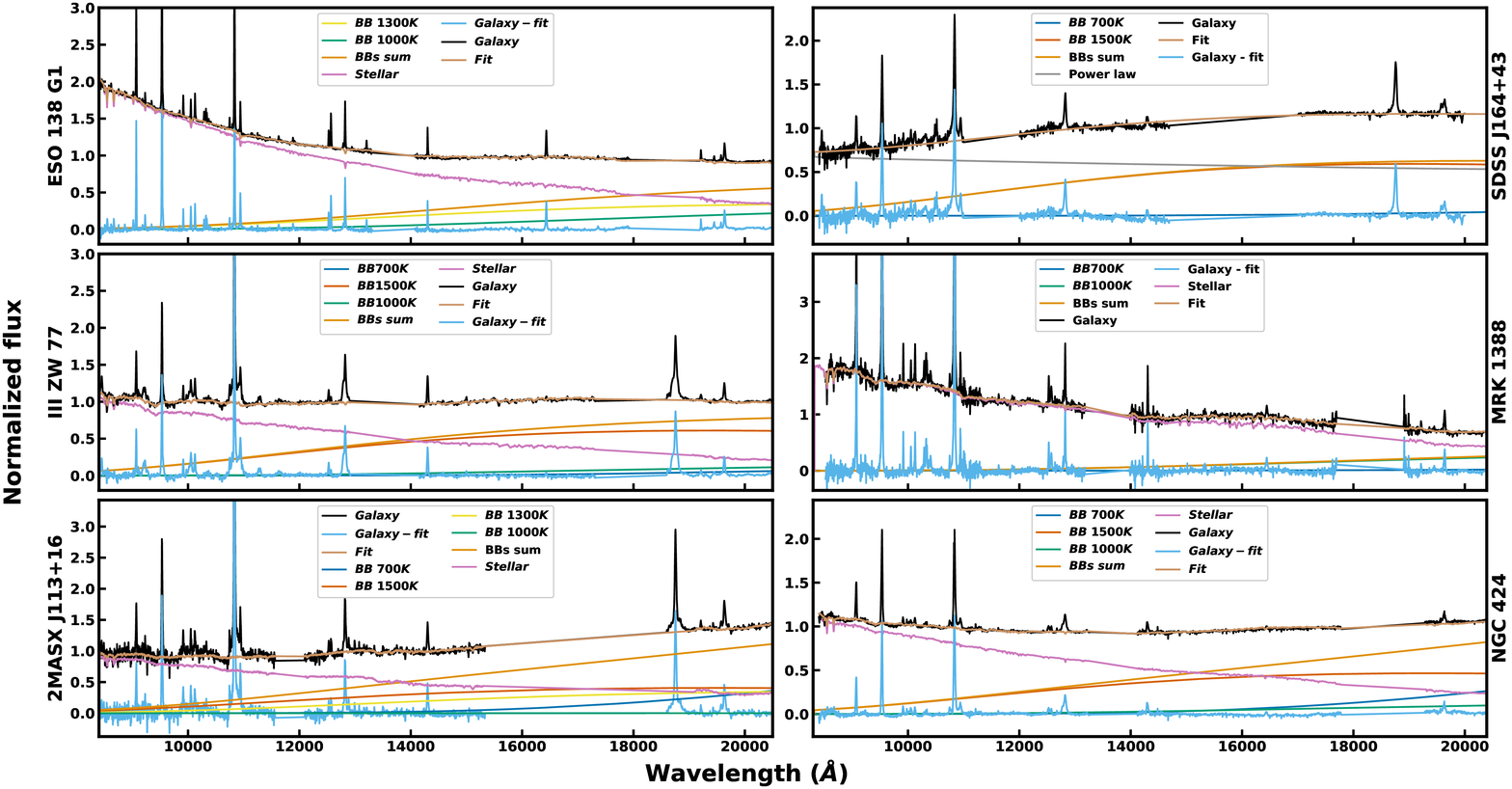}
    \caption{Fits of the continuum emission in the NIR. In all panels, the observed spectra is in black, the stellar template in magenta and the hot dust in orange. After subtraction of the continuum contribution, the residual consists of the nebular gas emission, shown in cyan.}\label{fig: ppxfFig}
    \end{center}
	
	\centering
\end{figure*}

\section{continuum emission analysis} \label{continuum}

In this section we describe the modelling of the continuum emission in the optical and NIR in order to remove the effects of both the underlying stellar population of the host galaxy and the continuum emission attributed to dust heated by the AGN. The first component is observed in both spectral regions while the second is restricted to the NIR.

\subsection{Optical continuum}

A rapid inspection to the optical spectra of the CLiF sample (see Fig.~1 of \citetalias{rose/2015a} and Fig.~\ref{fig:SOAR} of this manuscript) reveals the presence of strong absorption lines, evidencing the contribution of stellar population from the host galaxy to the observed integrated spectra. It is not of interest in this paper to study the stellar population, but to model and subtract this contribution to obtain suitable flux measurements for the gas in the central region. This procedure particularly impact the measurements of the Balmer decrement and the flux of lines that are partially absorbed by the stellar continuum.

In order to remove the stellar contribution, the Penalized Pixel-Fitting (pPXF) software developed by \cite{cappellari/2004} and updated by \cite{cappellari/2017} was employed. This routine uses maximum penalized likelihood estimation with the principle of searching for the best spectral fit by combining stellar spectra from a library of simple stellar populations. Here, the templates that were used to generate the pPXF fits are those of \cite{bruzual/2003}. Fig. \ref{fig: ppxfFig_opt} shows the fits carried out to the galaxy sample and the resulting nebular emission after subtraction of the stellar component. In addition, pPXF also estimates the stellar velocity dispersion, which will be employed in the determination of the mass of the black hole (see Sect.~\ref{sec:smbh}).

\subsection{The NIR continuum}\label{NIRcont}

Accepting the unified model to describe the different types of AGNs, the presence of a dusty torus leaves a spectroscopic signature that starts to show up  in the NIR. Dust absorbs a fraction of the optical/ultraviolet continuum from the central source and radiates it back at longer wavelengths, from 1~$\mu$m up to the far-infrared. 

Observational evidence corroborates models describing the NIR continuum in AGNs from 1 to 10 $\mu$m as being predominantly or entirely dominated by hot dust (\citealt{edelson/1986};  \citealt{barvainis/1987}). 

\citet{rodriguez/2006b} confirmed this scenario in the Seyfert\,1 AGN Mrk\,1239 by adjusting the excess of NIR continuum over the power-law extrapolated from the optical region, as due to dust emission. They employed a simple blackbody function at a temperature of 1200~K. This is close to the sublimation temperature of dust grains ($\sim$1500~K). 

Thus, to fit the continuum of our sample of CLiF AGN in the NIR, in addition to the stellar population templates, a function that represents the hot dust emission should be employed. For the former component, the IRTF library of stellar spectra \citet{rayner/2009} was used to fit the data. It is composed of 296 spectra in the wavelength range 0.8 to 5.0 $\mu$m, with a resolving power of approximately 2000. The SpeX/IRTF spectrograph was used for the observation of that library.

For the latter component, blackbody templates at the temperatures of 700~K, 1000~K, 1300~K and 1500~K were included following the procedure outlined in \citet{riffel/2009}. Dust emission at that temperature interval is expected for the innermost region of the torus and should leave a spectroscopic signature in the form of an excess of continuum emission over the power law, peaking at about 2$\mu$m.  The results are shown in Fig. \ref{fig: ppxfFig}. It can be seen that except in SDSS~J164+43, where no evidence of stellar population was found, the NIR continuum consists of the contribution of stellar light plus the hot dust component.  

\section{Measurement of the emission lines flux}\label{lines}

\begin{figure}

    \begin{center}
    \includegraphics[width=\columnwidth,height=10cm]{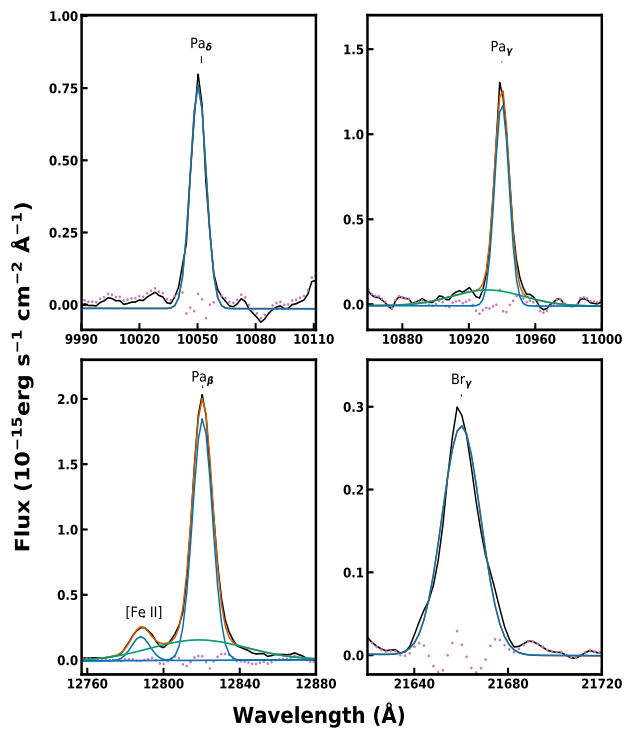}
    
    \caption{Multi-component Gaussian fit for the NIR of some H\,{\sc i} lines in the CLiF AGN ESO 138-G01. In each panel, observations are in black, narrow components in blue, broad components in green and total fit in red. The magenta dotted-line is the residuals after subtraction the total fit. It is possible to notice that the broad component is hidden for the less intense hydrogen lines.}\label{fig: eso_gau}
    
    \end{center}

	\centering
\end{figure}

\begin{figure}

    \begin{center}
    \includegraphics[width=\columnwidth,height=10cm]{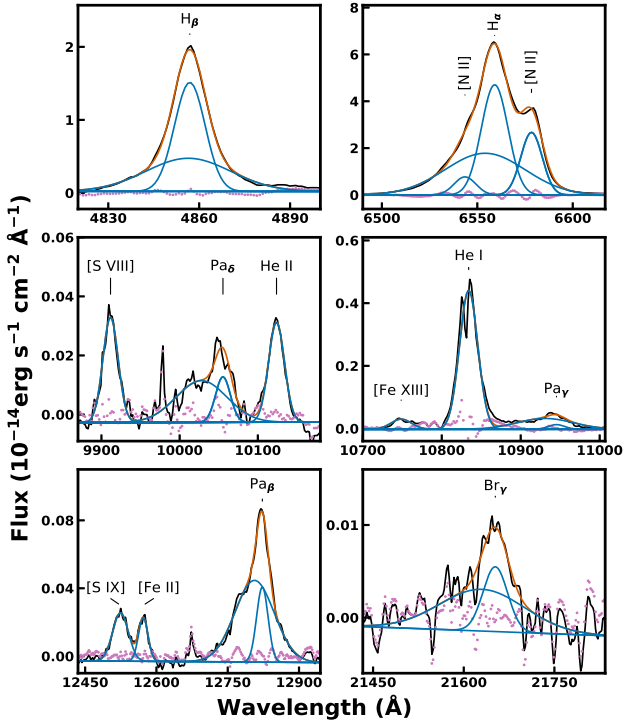}
    
    \caption{Multi-component Gaussian fit for the optical and NIR regions around the H\,{\sc i} lines in the CLiF AGN NGC\, 424. In each panel, observations are in black, individual components in blue, total fit in red. The magenta dotted-line is the residuals after subtraction the total fit.}\label{fig: ngc 424_gau}
    
    \end{center}

	\centering
\end{figure}

\begin{figure}

    \begin{center}
    \includegraphics[width=\columnwidth,height=6cm]{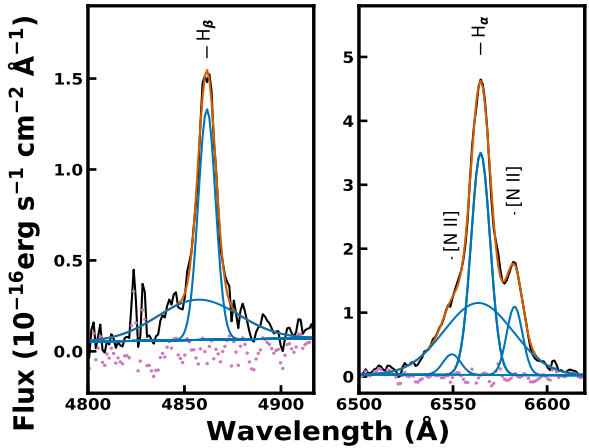}
    
    \caption{Same as Fig.~\ref{fig: ngc 424_gau} for H$\beta$ (left panel) and H$\alpha$ (right panel) observed in SDSS J164126.91 + 432121.\label{fig: j164_gau}}
    
    \end{center}

	\centering
\end{figure}

\begin{figure}

    \begin{center}
    \includegraphics[width=\columnwidth,height=6cm]{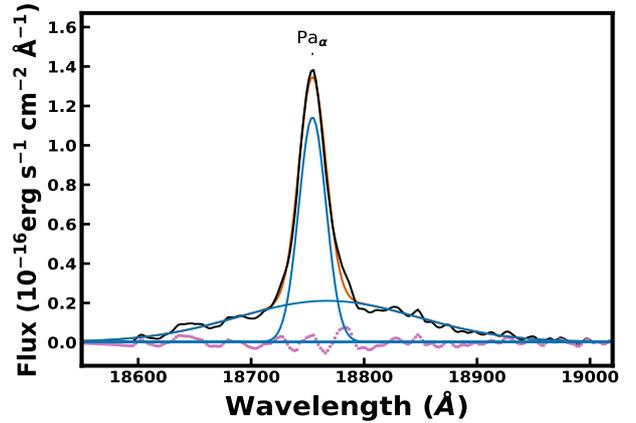}
    
    \caption{Same as Fig.~\ref{fig: ngc 424_gau} for Pa$\alpha$ in 2MASX J113111.05+162739.}\label{fig: 2masx_gau}
    
    \end{center}

	\centering
\end{figure}

After subtraction of the stellar and hot dust contribution, the resulting spectra consist of emission lines produced by the gas from the nuclear and circumnuclear regions of the galaxies. In order to measure the flux of the lines, we modelled the observed profiles with a suitable function that best represents them and then integrated the flux under that function. To this purpose we employ the LINER routine (\citealt{pogge/1993}).  This software performs a least-square fit of a model line profile (Gaussian, Lorentzian, or Voight functions) to a given line or set of blended lines to determine the flux, peak position and FWHM of the individual components. Typically, one or two Gaussian components were necessary to represent the NLR lines. For permitted lines with a clear broad component associated with the BLR, a third Gaussian function was added. Figures~\ref{fig: eso_gau} to ~\ref{fig: mrk1388}. show examples of the deblending procedure carried out to the most important emission lines in the sample. Tables A1 to B6 list the peak position (column 2), ionization potential (IP, column 3), FWHM (column 4) and integrated flux (column 5) measured for all lines identified in each CLiF AGN both in the optical and the NIR.

\subsection{Detection of hidden broad lines in Type~II CLiF AGN} \label{sect:reclass}

\cite{rose/2015b} employed the colour index [W2-W4] derived from WISE to verify the inclination angle of the torus relative to the observer. From their analysis, they found that the [W2-W4] index in CLiF AGN was intermediate between objects classified as Type~I and Type ~II AGN. This result was consistent with the hypothesis that the torus in CLiFs are observed at intermediate inclination angles. However, as can be seen in the last column of Table \ref{tab:dados}, only one out of the seven CLiF AGNs is a genuine Type~II object. In most targets, broad components in the permitted lines were observed, even in those objects  previously classified as Type~II AGN.  This result points out that the majority of CLiF AGN are indeed Type~I AGN.

\begin{figure}

    \begin{center}
    \includegraphics[width=\columnwidth,height=6cm]{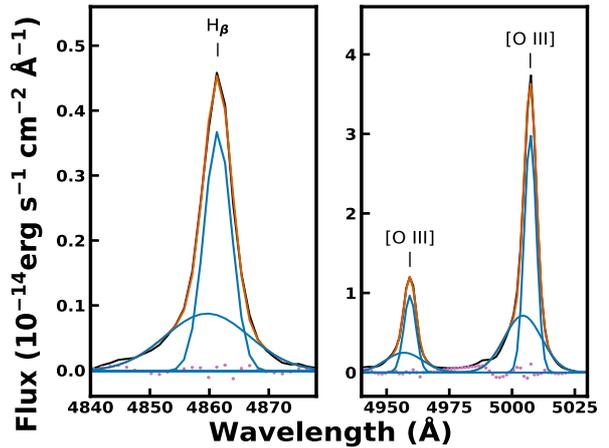}
    
    \caption{Same as Figure~\ref{fig: ngc 424_gau} for H$\beta$ (left panel) and \OIII\ (right panel) observed in MRK 1388.}\label{fig: mrk1388}
    
    \end{center}

	\centering
\end{figure}

The classification of MRK~1388 is ambiguous. It displays a broad component in the Balmer lines with FWHM of 1040 km~s$^{-1}$. This would tentatively classify it as a narrow line Seyfert~1 (NLS1). However, a similar broad component is also detected (in blue-shift) in the \OIII\ lines (see Fig. \ref{fig: mrk1388}), which suggests the presence of an outflow rather than a genuine BLR. Previously, \cite{doi/2015} classified MRK~1388 as a NLS1 because of the strong featureless continuum observed in that source although the optical spectrum does not display evidence of broad components in the permitted lines. He suggested that this galaxy has a heavily obscured BLR. 

On our NIR spectrum of MRK~1388, a similar broad component as the one detected in \OIII\ is also found in [S\,{\sc iii}] (see Table~\ref{tab:mrkinfra_tabelao}). It suggests that this object, indeed, displays evidence of an outflow. Notice that no broad NIR \ion{H}{i} line was found. As the optical and NIR spectra are not contemporaneous, it is possible that variability between both set of data can explain the lack of a broad \ion{H}{i} component in the NIR. Reports of such extreme variability is frequently found in the literature and is exemplified by the so-called changing-look AGNs. These objects
show extreme changes of emission-line intensities, with
sometimes almost complete disappearance and reappearance of
the broad component (see, e.g., \citealt{lyutyj/1984}; \citealt{kollatschny/1985}; \citealt{denney/2014}; \citealt{oknyansky/2019}). Although variability of coronal lines has been detected before in some sources (e.g., \citealt{landt/2015} and \citealt{landt/2015b}), the results show that they vary differently as the permitted lines.

In the light of the evidence presented above, we keep the classification of MRK\,1388 as a NLS1 galaxy.

In summary, the optical and NIR spectroscopy presented here reveals that just one (SDSS J124+44) out of the seven CLiF AGN display spectral characteristics typical of Type~II objects.  This is at odds with the model presented in \citet{glidden/2016} to explain CLiF AGN,  as it applies only to sources optically classified as Type~II.

\subsection{Extinction of the NLR in CLiFs}

In their optical study of CLiF AGN, \citetalias{rose/2015a} found out that the deviations from the intrinsic Balmer decrement H$\alpha$/H$\beta$ = 3.1 derived for their sample was attributed to high-density gas in the region where the H$\alpha$ line is formed. However, the lack of subtraction of the stellar continuum may hide part of the H$\beta$ emission. This is because part of the line is within the absorption dip produced by the stellar population. As a result, the line ratio H$\alpha$/H$\beta$ is overestimated relative to their theoretical value even in the absence of dust extinction.  

In this work, thanks to the increase of wavelength coverage by including NIR spectra, we expanded the number of diagnostic lines sensitive to extinction, allowing us to investigate this interesting issue in more detail. 
For this purpose, we determined the extinction affecting the gas by comparing observed to predicted Case-B recombination emission-line flux ratios, assuming the \cite{cardelli/1989} extinction law for R$ _ {V}$ = 3.1. To obtain E(B-V), we employed the fluxes F$ _ {\textup{pa}\beta} $, F$ _ {\textup{pa}\gamma}$, F$ _ {\textup{pa}\delta} $ and F $ _ {\textup{Br}\gamma} $, which represent respectively the Hydrogen line fluxes Pa$\beta$, Pa$\gamma$, Pa$\delta$ and Br$\gamma$.  The expressions of \citet{rodriguez/2016} were employed to this purpose. In addition to these NIR \ion{H}{i} lines, we also determined the extinction by means of the Balmer decrement H$\alpha/$H$\beta$ in the optical region.

Finally, we also employed [\ion{Fe}{ii}] lines in the NIR to determine E(B-V). This is because the transitions leading to the \FeII\ $ 1.257 \mu$m and 1.644 $\mu$m lines originate from the same upper level. In the optical thin case, the ratio of their intrinsic line fluxes is simply given by the ratio of the corresponding spontaneous emission coefficients, which is constant and in principle known. Therefore, they are suitable diagnostics of extinction caused by dust due to the large wavelength separation between the lines. The intrinsic value of the ratio \FeII\ $1.257 \mu$m / 1.644 $\mu$m is estimated to be 1.25 with an uncertainty of 20 \% (\citealt{bautista/2015}). In order to determine E(B-V) using this ratio we employed the expression derived by \citet{rodriguez/2016}.

\citet{riffel/2006} and \citet{rodriguez-ardila/2017b} had already shown that the ratio of the \FeII lines 1.257 $\mu$m / 1.644 $\mu$m is in accordance with the extinction obtained using the Pa$\beta$/Br$\gamma$ ratio and could be reliably applied to Seyfert 1 and 2 galaxies.

The values of E(B-V) obtained using the 5 different dust sensitive line ratios are listed in columns 2 to 6 of Table \ref{tab:ext_indics}. Column 7 of the same table shows the final extinction for each galaxy, determined after averaging out the individual entries of the different indicators.

The overall very good match between the $E(B-V)$ found from different
indicators for the same galaxy indicates the consistency of our approach. Although some discrepancies are noted at some specific line ratios, usually differences of less than $\sim$0.3 mag between the minimum and maximum values of $E(B-V)$ were observed. We found that CLiF display extinction values similar to those found in samples of non-CLiF AGN. For instance, Malkan et al. (2017) measured an average $E(B-V)$ of 0.49 $\pm$ 0.35 for
type 1 and 0.52 $\pm$ 0.26 for type 2 Seyferts in their study of emission-line properties of Seyfert galaxies. They used a sample of 81 Seyfert~1 and 104
Seyfert~2 galaxies that comprise nearly all of the IRAS 12~$\mu$m AGN sample. Thus, in terms of intrinsic extinction, CLiF AGN are rather similar to the non-CLiF counterparts.

\begin{table*}
\centering
  \caption{E(B-V) (in mag) obtained for line ratios}
\begin{tabular}{lcccccc}
\hline
\multicolumn{1}{c}{Galaxy}      & E(B-V)$_{Pa_{\beta }/Br_{\gamma }}$ & E(B-V)$_{Pa_{\gamma }/Br_{\gamma }}$ & E(B-V)$_{Pa_{\delta }/Br_{\gamma }}$ & E(B-V)$_{[Fe II]}$    & E(B-V)$_{H_{\alpha  }/H_{\beta  }}$                              & \textless{}E(B-V)\textgreater{} \\ \hline
ESO 138 G1   & 0.76  $\pm$ 0.12             & 0.47  $\pm$ 0.04                    & 0.46  $\pm$ 0.03                     & 0.68  $\pm$ 0.04      & 0.46  $\pm$ 0.01                                                 & 0.57  $\pm$ 0.05                \\
SDSS J164+43 & -                                  & -                                   & -                                    & -                     & 0.22 $\pm$ 0.01                                                  & 0.22 $\pm$ 0.01                 \\
III ZW 77    & 0.23 $\pm$ 0.06                    & 0.52 $\pm$ 0.13                     & 0.38 $\pm$ 0.06                      & 1.16 $\pm$ 0.67$^{*}$ & 0.17 $\pm$ 0.01                                                  & 0.33 $\pm$ 0.05                 \\
MRK 1388     & 0.32 $\pm$ 0.26   & 0.10 $\pm$ 0.05    & 0.16 $\pm$ 0.06  & 0.00 $\pm$ 0.00 & 0.00 $\pm$ 0.00  & 0.12 $\pm$ 0.07                 \\
SDSS J124+44 & -                                  & -                                   & -                                    & -                     & 0.27 $\pm$ 0.01                                                  & 0.27 $\pm$ 0.01                 \\
NGC 424      & 0.27 $\pm$ 0.08                    & 0.29 $\pm$ 0.06                     & 0.13 $\pm$ 0.03                      & 0.02 $\pm$ 0.01 & 0.45 $\pm$ 0.03                                                  & 0.23 $\pm$ 0.04                 \\ \hline
\multicolumn{7}{l}{$^{*}$value not taken into account on average}   
\end{tabular}
\label{tab:ext_indics}
\end{table*}

Because all dust-sensitive NIR flux ratios employed here point towards small to moderate amounts of extinction, we conclude that the CLiF AGN are affected, to some extend, by reddening in the nuclear region. Moreover, our values of the H$\alpha$/H$\beta$ ratios were, in general, smaller than those of \citetalias{rose/2015a}  because of the subtraction of the stellar component. Indeed, the presence of the stellar population should affect the Balmer decrements, as the apertures employed to extract the spectra cover region sizes of the order of kiloparsecs.

This effect was already expected as the continuum emission displayed  by the galaxies showed strong evidence of an underlying stellar population. In this scenario, H$\beta$ tends to be more absorbed than H$\alpha$, increasing artificially the flux ratio H$\alpha$/H$\beta$.
It is also worth to mention that the values of E(B-V) found here are low enough so that the H$\gamma$/H$\beta$ ratio is not significantly affected. It tends to be close to the theoretical value as the separation in wavelength of the above two lines is small.

In Table \ref{tab:red&decre} we list the H$\alpha$/H$\beta$ ratios obtained by \citetalias{rose/2015a} and the ratios obtained by us after removing the stellar population using pPXF. Because of the consistency of our results using a larger number of diagnostic lines, throughout this work we consider that deviations from the intrinsic \ion{H}{i} or \FeII\ $1.257 \mu$m / 1.644 $\mu$m ratios are due to extinction rather than to density effects. Thus, the emission-line fluxes for each galaxy were corrected by the corresponding E(B-V) listed in the last column of Table \ref{tab:ext_indics}.  

%%%%%%%%%%%%%%%%%%%%%%%%%%%%%%%

\begin{table*}
\centering
\caption{Comparison between values measured by \protect\citetalias{rose/2015a} (REF) and after the star population subtraction in our analysis (Pop. Sub.).}
\begin{tabular}{lccccc}
\hline
\multicolumn{1}{c}{\textbf{Galaxy}} & <E(B-V)>      & H$\alpha$/H$\beta_{REF}$ & H$\alpha$/H$\beta_{Pop. \ Sub.}$ & H$\gamma$/H$\beta_{REF}$ & H$\gamma$/H$\beta_{Pop. \ Sub.}$ \\ \hline
ESO 138 G1                          & 0.49$\pm$0.05 & 4.25$\pm$0.10            & 4.92$\pm$0.05                    & 0.48$\pm$0.02            & 0.33$\pm$0.01                    \\
SDSS J164+43                        & 0.22$\pm$0.01 & 5.38$\pm$0.19            & 3.72$\pm$0.43                    & 0.40$\pm$0.03            & 0.37$\pm$0.06                    \\
III ZW 77                           & 0.33$\pm$0.05 & 4.15$\pm$0.13            & 3.49$\pm$0.39                    & 0.34$\pm$0.04            & 0.47$\pm$0.09                    \\
MRK 1388                            & 0.19$\pm$0.05 & 3.95$\pm$0.12            & 2.71$\pm$0.90                    & 0.49$\pm$0.03            & 0.53$\pm$0.03                    \\
SDSS J124+44                        & 0.27$\pm$0.01 & 6.36$\pm$0.21            & 3.90$\pm$0.35                    & 0.47$\pm$0.04            & 0.45$\pm$0.11                    \\
2MASX J113+16                       & -             & 5.00$\pm$0.17            & -                                & 5.00$\pm$0.17            & -                                \\
NGC 424                             & 0.29$\pm$0.03 & 5.77$\pm$1.00            & 4.85$\pm$0.37                    & 0.48$\pm$0.05            & 0.42$\pm$0.04                    \\ \hline
\end{tabular}
\label{tab:red&decre}
\end{table*}

%%%%%%%%%%%%%%%%%%%%%%%%%%%%%%%%%%%%%%%%%%

\subsection{Spectroscopic properties of CLiF AGNs and comparison with other AGN}

One of the most important characteristics of CLiF AGN is the strength of the coronal lines, quantified by means of the emission-line flux ratios [\ion{Fe}{vii}]~$\lambda$6087/H$\beta$, [\ion{Fe}{x}]~$\lambda$6374/H$\beta$ and [\ion{Ne}{v}]~$\lambda$3425/H$\beta$. In CLiF AGN the above ratios should be larger than 0.25, 0.2 and 1, respectively. In all cases,  H$\beta$ corresponds to the flux of the narrow component if the AGN is of Type~I. 

As it was shown in the previous section, H$\beta$ may be strongly diluted by the presence of an underlying stellar population. Therefore, it is important to verify if the values of the above three line ratios still hold in CLiF AGNs after removing the stellar continuum. Moreover, if dust is present, correction by extinction will decrease [\ion{Fe}{vii}]~$\lambda$6087/H$\beta$ and [\ion{Fe}{x}]~$\lambda$6374/H$\beta$ and increase [\ion{Ne}{v}]~$\lambda$3425/H$\beta$.  

\begin{table}
\centering
\caption{Most relevant optical line ratios that define a CLiF after subtraction of the stellar population and correction by extinction.}
\begin{tabular}{lccc}
\hline
Galaxy & [\ion{Ne}{v}]/H$\beta$ & [\ion{Fe}{vii}]/H$\beta$ & [\ion{Fe}{x}]/H$\beta$  \\
\hline
ESO\,138\,G1  &	0.58$\pm$0.05 &	 0.21$\pm$0.01 & 0.07$\pm$0.01  \\
SDSS\,J164+43 &	5.13$\pm$0.38 &	 0.38$\pm$0.07 & 0.11$\pm$0.01 \\
III\,ZW\,77	  & ...	 	      &  0.16$\pm$0.02 & 0.08$\pm$0.01  \\
MRK\,1388	  & ...	 	      &  0.33$\pm$0.01 & 0.14$\pm$0.01  \\
SDSS\,J124+44 &	...	 	      &  0.54$\pm$0.07 & 0.18$\pm$0.04  \\
NGC\,424	  & 1.96$\pm$0.09 &	 0.27$\pm$0.04 & 0.08$\pm$0.02  \\
\hline
\end{tabular}
\label{tab:newrat}
\end{table}

Table~\ref{tab:newrat} lists the line ratios between the most prominent coronal lines, normalized to H$\beta$ narrow component after subtraction of the stellar population and corrected by extinction. The flux values employed are listed in Tables~\ref{tab:esoopt_tabelao} to~\ref{tab:ngcopt_tabelao}.  overall, the  derived ratios are close (within the uncertainties) to the ones defined by \citet{rose/2015a}.  It means that the limits established by these authors can be considered as guidelines to select sources with similar characteristics. However, in order to study the physical properties of these objects, the removal the stellar continuum or correction by reddening must be first carried out.

\begin{figure}

    \begin{center}
   \includegraphics[width=\columnwidth]{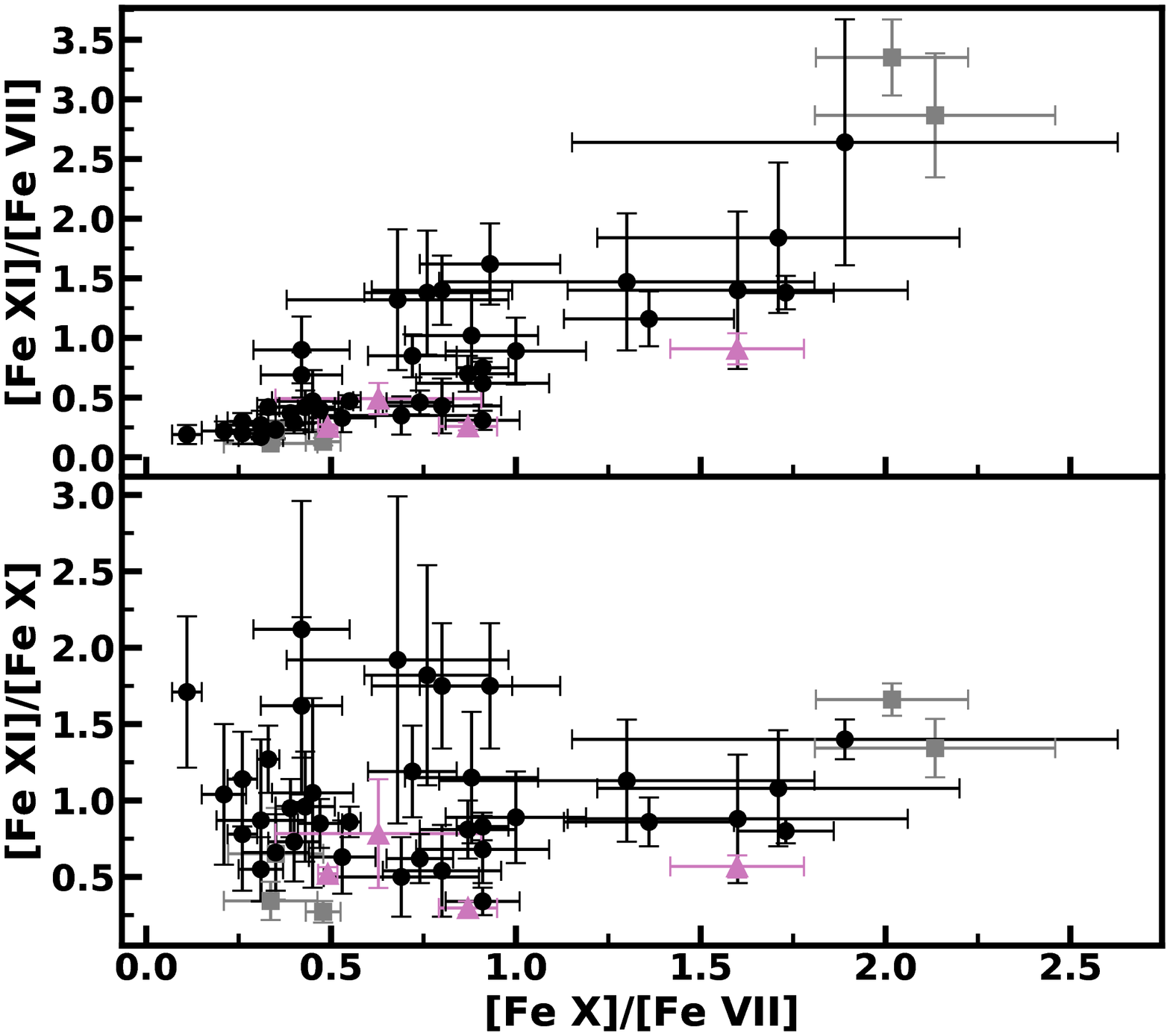}
    
    \caption{Observed \FeXI/\FeVII \ (top panel) and \FeXI/\FeX \ (bottom panel) vs \FeX/\FeVII ratios for the CLiF AGNs (magenta points). For comparison, the grey points are taken from \protect\cite{rodriguez/2006} and black points from \protect\cite{gelbord/2009}. }\label{fig:optcoronalratio1}
    \end{center}

	\centering
\end{figure}

At this point, is interesting to examine the relative strength of CLs in CLiF AGN and compare the results with values derived for other AGN in the literature. To this purpose, we calculated the line flux ratios [\ion{Fe}{x}]~$\lambda6374$/[\ion{Fe}{vii}]~$\lambda6087$, [\ion{Fe}{xi}]~$\lambda7890$/[\ion{Fe}{x}]~$\lambda6374$ and [\ion{Fe}{xi}]~$\lambda7890$/[\ion{Fe}{vii}]~$\lambda6087$ in our CLiF sample. Moreover, data from \cite{rodriguez/2006} and \cite{gelbord/2009} were used for comparison purposes. We will refer hereafter to the data taken from the literature as the comparison sample. For the present study, we included only galaxies that had a secure detection in the three iron emission lines above. This is because in some cases, upper limits are driven by spurious residuals left after telluric correction or bad sky subtraction. Thus, to warrant meaningful values in the derived ratios, only fluxes detected at 3$\sigma$ were employed. The results are shown in Fig.~\ref{fig:optcoronalratio1}. It is important to highlight that \cite{gelbord/2009} selected the 63 AGN of their sample based on the 3$\sigma$ detection of the \FeX \ $\lambda$6374 line. Of these, 55 spectra included the spectral region containing \FeXI \ although it was detected (3$\sigma$) in only 37. Therefore, \FeXI \ was observed in 65\% of the sample. In our sample of CLiF AGNs, four objects included the region of the \FeXI \ and in all of them  that line was detected.  

An inspection to Fig.~\ref{fig:optcoronalratio1} shows that CLiF (full magenta triangles) and non-CLiF AGN (full black circles and full gray squares) display similar values in the line ratio [\ion{Fe}{x}]/[\ion{Fe}{vii}], plotted in the x-axis of both panels. Likewise, the two ratios plotted in the y-axis, [\ion{Fe}{xi}]/[\ion{Fe}{x}] (bottom panel) and [\ion{Fe}{xi}]/[\ion{Fe}{vii}] (upper panel), are comparable in CLiFs and non-CLiF AGN. Under the assumption that the ratios depicted in the y-axis measure the ionization state of the CL gas, CLiF AGN do not represent an extreme case within the AGN class.  Therefore, we can safely state that when line flux ratios between optical CLs are considered, CLiFs are not a separate class of active galactic nuclei.

\begin{figure}

    \begin{center}
    \includegraphics[width=\columnwidth]{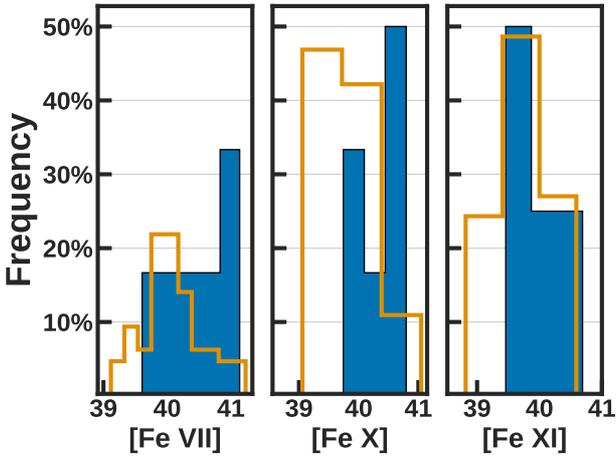}
    
     \caption{Histogram of luminosity values (log erg s$^-1$) for \FeVII, \FeX \ and \FeXI \ for CLiF AGNs (in blue). For comparison, measures obtained by \protect\cite{gelbord/2009} (in orange) were used. }\label{fig:lum_iron}
    \end{center}

	\centering
\end{figure}

We also derived the luminosity of the most prominent optical CLs in CLiF AGN and compared these values to those found in the comparison sample.  The results are shown in Fig.~\ref{fig:lum_iron}, where the luminosity distribution of [\ion{Fe}{vii}] (left panel), [\ion{Fe}{x}] (central panel) and [\ion{Fe}{xi}] (right panel) is presented. The shaded blue histogram is for CLiF while the orange histogram represents non-CLiF AGN. It can be seen that all three lines are distributed in a narrow interval of luminosity (10$^{39-41}$~erg\,s$^{-1}$) but CLiF AGN tend to occupy the high-end of the three distributions. 

\begin{figure}

    \begin{center}
    \includegraphics[width=\columnwidth, height=5cm]{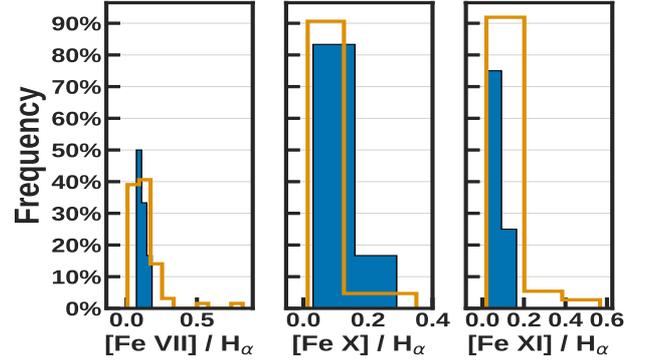}
    
     \caption{histograms of the observed \FeVII/H$_{\alpha}$, \FeX/H$_{\alpha}$ and \FeXI/H$_{\alpha}$ for the CLiF AGNs (in blue). For comparison, measures obtained by \protect\cite{gelbord/2009} (in orange) were used. It is possible to verify no great distinction between our sample and non-CLiF AGNs. }\label{fig:razoesH_Ferro}
    \end{center}

	\centering
\end{figure}

Moreover, we study the distribution of the ratios between CLs and \ion{H}{i} lines for the three iron lines above in CLiF and non-CLiF AGN. The main goal here is to compare if the values observed in the former are indeed extreme within the AGN class. The results, presented in Fig.~\ref{fig:razoesH_Ferro}, show precisely the opposite. The values of the ratios [\ion{Fe}{vii}]/H$\beta$, [\ion{Fe}{x}]/H$\alpha$ and 
[\ion{Fe}{xi}]/H$\alpha$ found in CLiFs are just part of a wider interval of values displayed by the comparison sample.

The availability of NIR spectra allowed us to study, for the first time in the literature, the spectroscopic properties of CLIF in this spectral region. Our main goal is to explore how these sources behave in the NIR when compared to non-CLiF AGN. 

A close examination to Fig.s~\ref{fig:gemini} to~\ref{fig:blanco} reveals that in the NIR, optically classified CLiF display strong CLs of Aluminium, Calcium, Iron, Silicon and Sulfur. A comparison of our sample spectra to classical AGN presented in \cite{rodriguez/2011} shows the same ionization species. However, two important facts deserve mentioning here. 

First, each CLiF AGN of this work displays all CLs already identified in previous samples in the interval 0.8$-$2.4$\mu$m.  In contrast, \citet{rodriguez/2011} reported that in only 3 (NGC\,1068, NGC\,4051 and Ark\,564) out of 54 AGN, all CLs considered here were simultaneously detected. Moreover, they report that the four most prominent CLs in the NIR, \SiVI, \SiX, \SVIII\ and \SIX, showed up all together in just 30\% of their sample. In a later work with a larger number of targets (104 AGN), \cite{lamperti/2017} found that in just 18\% of their sample, more than two coronal lines were detected in the same spectrum. We ruled out sensitivity issues for the non-detection of CLs in most of the sources employed for comparison. However, in some cases, the spectral regions containing the lines are out of the filter bandpasses or severely affected by atmospheric transmission. In this work, the above four lines appeared in all six CLiF AGN. 
Second, the [\ion{Fe}{xiii}] doublet at 1.074, 1.0798~$\mu$m was detected in all CLiF here while it was identified in just 15\% of the non-CLiF sample of \cite{rodriguez/2011} and in 5\% of objects in \cite{lamperti/2017}.

\begin{figure}

    \begin{center}
    \includegraphics[width=\columnwidth]{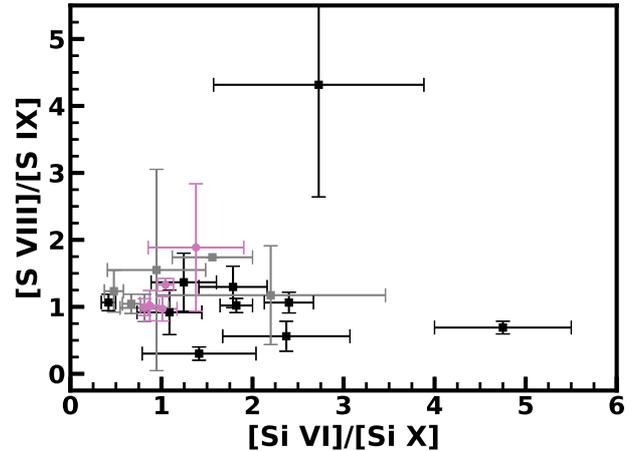}
    
     \caption{Observed \SiVI/\SiX \ vs \SVIII/\SIX\ ratios for the CLiF AGN (magenta points). Black and grey points are for Type~I and~II AGN, respectively, taken from \protect\cite{rodriguez/2011}. }\label{fig:beto+CLiF}
    \end{center}

	\centering
\end{figure}

Following the analysis carried out in the optical, we compare the relative intensity of CLs in CLiF  to that of non-CLiF AGN in the NIR. To this purpose, we determined the line ratios \SiVI / \SiX \ and \SVIII / \SIX from the fluxes listed in Tables~\ref{tab:esoinfra_tabelao} to~\ref{tab:ngcinfra_tabelao} and compare them to the values reported in \cite{rodriguez/2011}.Only 3$\sigma$ detections were employed. Note that  each ratio is independent of the gas metallicity. The results, shown in Fig. \ref{fig:beto+CLiF}, suggest that CLiF have little scatter in both ratios, being concentrated mostly in the region with values of \SiVI / \SiX \ and \SVIII / \SIX\, of $\sim$1. The scatter observed in non-CLiF is indeed larger, with most points distributed in the intervals 0.3$-$2.2 and 0.2$-$2 for \SiVI / \SiX \ and \SVIII / \SIX, respectively. From this comparison, we verify that regarding coronal line ratios, no clear distinction between CLiF and non-CLiF AGNs is observed.

\begin{figure}

    \begin{center}
    \includegraphics[width=\columnwidth]{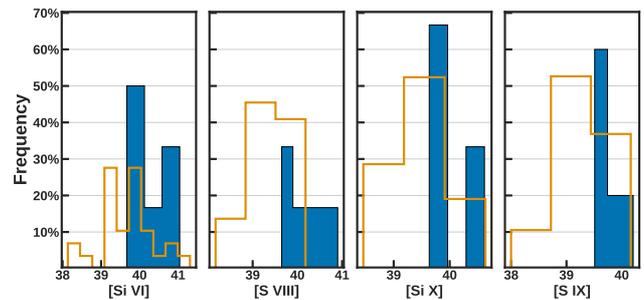}
    
     \caption{Histogram of the line luminosity distribution (log erg s$^{-1}$) for \SiVI, \SVII, \SiX \ and \SIX \ for CLiF AGNs (in blue). For comparison, data obtained by \protect\cite{rodriguez/2011} for non-CLiF AGN (orange) were used. }\label{fig:lum}
    \end{center}

	\centering
\end{figure}

We also derive the luminosity of the above four NIR coronal lines and compare the distribution of values to that of non-CLiF AGN. The results, shown in Fig.~\ref{fig:lum} confirm the trend already observed in the optical region. The values of CL luminosity in CLiFs are preferentially located in the upper end of the distribution shown by non-CLiF AGN. Indeed, CLiF occupy a very narrow range of values in luminosity: 10$^{40-41}$~erg\,s$^{-1}$ for \SiVI, \SVIII\ and \SiX\ and $\sim$10$^{39-40}$~erg\,s$^{-1}$ for \SIX. Non-CLiFs stretch over a luminosity range of $2-3$ dex, between 10$^{39}$ and 10$^{39}$~erg\,s$^{-1}$. 

\begin{figure}

    \begin{center}
    \includegraphics[width=\columnwidth, height=5cm]{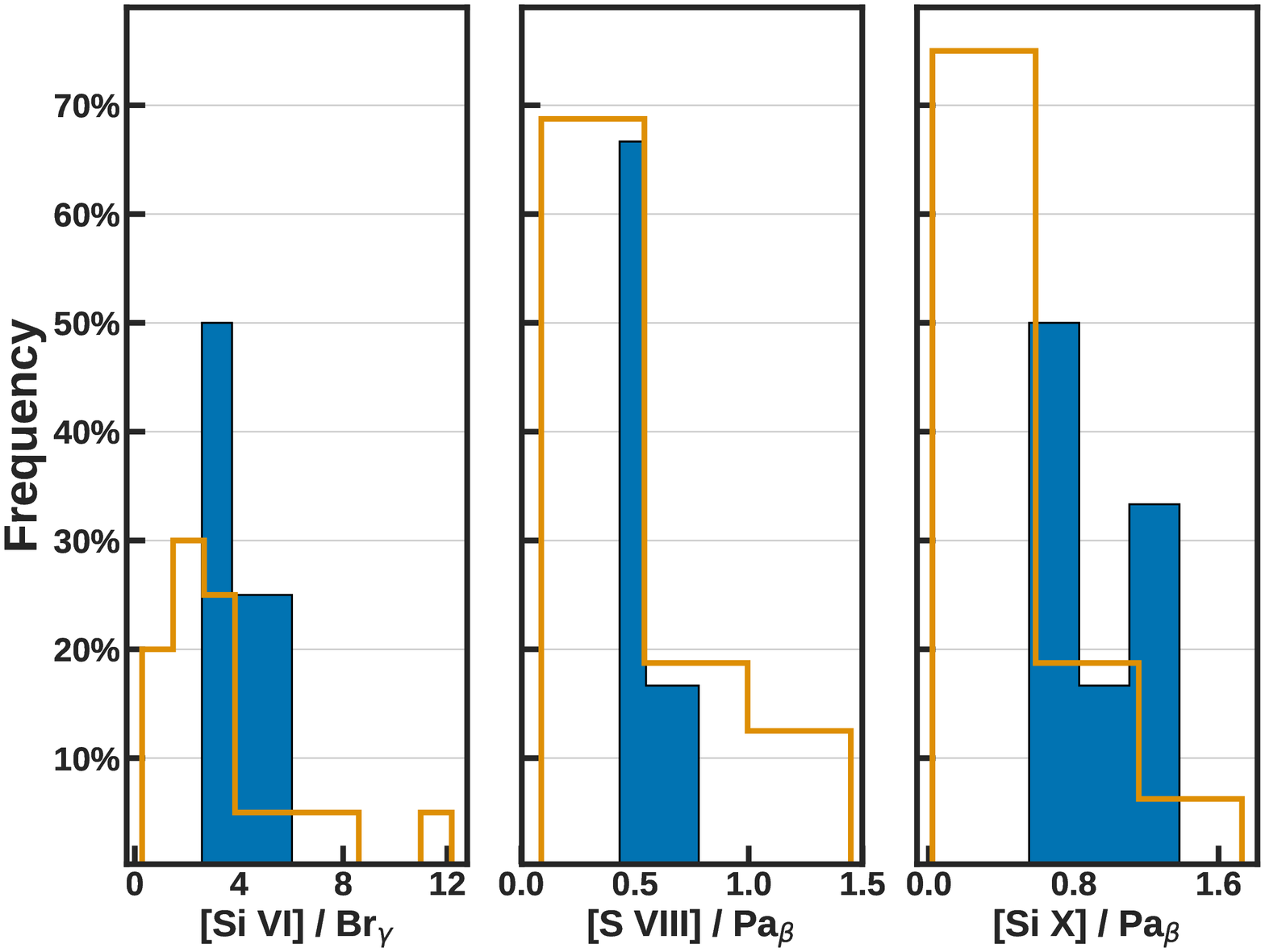}
    
     \caption{histograms of the observed \SiVI/Br$_{\gamma}$, \SVII/Pa$_{\beta}$ and \SiX/Pa$_{\beta}$ ratios for the CLiF AGNs (in blue) and non-CLiF AGN (orange). Data for the later sample are from \protect\cite{rodriguez/2011}.  }\label{fig:razoes}
    \end{center}

	\centering
\end{figure}

Finally, we also constructed the emission-line flux ratios \SiVI/Br$\gamma$, \SVIII/Pa$\beta$ and \SiX/Pa$\beta$ to assess whether CLiF tend to display larger values in these ratios than in non-CLiF.  Fig.~\ref{fig:razoes} shows the corresponding histograms. It can be easily seen that \SiVI\ and \SiX\ tend to be  brighter relative to low-ionization lines in CLiF.

The results presented in the preceding paragraphs show that regarding the relative strength of the most important coronal lines in the optical and NIR, CLiFs are not distinguished among other AGNs.  However, high-ionization lines in CLiFs are usually confined to the upper-half of the luminosity distribution presented by AGNs in general. This tendency is also observed when ratios between CLs and \ion{H}{i} lines are examined. CLiFs tend to occupy the upper bins of the distributions when confronted to the general AGN population. We are aware, though, that the small number of objects that comprises our CLiF sample precludes into a more conclusive statement. However, it is clear from our analysis that CLiFs, by no means, represent a separate class of AGN.

One possible scenario to enhance the high-ionization spectrum relative to that of lower ionization is to consider that the NLR is dominated by matter-bounded clouds. This hypothesis was proposed by \cite{viegas/1992} and \cite{binette/1996} while trying to understand the nuclear and extended NLR emission of a sample of AGN. They showed that the inclusion of photoionized
matter-bounded (MB) clouds of sufficiently high excitation (high ionization parameter) successfully reproduce the high excitation lines. The low excitation lines were accounted for by a population of low ionization parameter, ionization-bounded (IB) clouds exposed to the ionizing radiation spectrum that leaks from the MB clouds. Varying the relative proportion of the two types of clouds was found to have a similar effect on the combined emission-line spectrum to varying the ionization parameter in a traditional ionization parameter sequence.

This model was later applied by \cite{binette/1997} to the Circinus Galaxy, a  Type~II AGN widely known for its outstanding coronal emission-line spectrum. In order to explain the CLiF AGN under this scenario, the number of MB-clouds should largely exceed that of IB clouds. Thus, the resulting spectrum is dominated by that emitted by the MB component. Although making a detail modelling of the CLiF AGN is out of the scope of this paper, we will examine in the next section the physical properties of NLR gas. This will allow us to further constrain this scenario and if it can be applied to our sample of AGN. 

Another mechanism that can enhance the high-ionization lines is shocks, driven by the interaction of a radio jet and the NLR gas or by nuclear outflowing material. The combined effect of nuclear photoionization and shocks strongly influence the strength of the CLs (\citealt{contini/2001}). This hypothesis was successfully applied to a local sample of AGN in \citet{rodriguez/2006}, where evidence of high-ionization outflows were detected.  According to \citet{contini/2001} models, shock velocities of V$\geq$500~km\,s$^{-1}$ are necessary in order to produce very high ionization lines such as [\ion{Fe}{x}] or [\ion{Si}{x}]. Although the gas kinematics will be discussed in Section~\ref{kinematics}, we anticipate that evidence of CL-driven shocks in the CLiF sample exist but cannot be applied to all targets studied here. For instance, broad components in the forbidden lines are observed in some of the sources, mainly in the [\ion{S}{iii}] line. In NGC\,424, broad features are also observed in the CLs. However, in ESO\,138-G001, Mrk\,1388 and J124134.25+442639.2 no evidence of broad CL were found.
Although we do not discard the presence of shocks as an additional mechanism to enhance the CL emission, it is not be the main driver of the rich coronal line spectrum that characterises CLiF AGN.

At this point, we also highlight the identification of \SXII\ 7611\text{\AA}, whose ionization potential of 504.8~eV is the highest among all NIR and optical coronal lines examined here. \cite{osterbrock/1981} had already noted the presence of an unidentified feature at 7613.1$\pm$2.8\,\text{\AA} in III\,Zw\,77 but he did not associated it with any particular emission-line. \cite{kraemer/2000} were the first to attribute the line at 7611\,\text{\AA} detected in NGC\,1068  to \SXII. Here, we confirm the identification of \SXII\ in III\,Zw\,77 and MRK\,1388. In the remaining CLiF AGN, the region containing that feature is outside of the spectral coverage of the data. Note that, to the best of our knowledge, \SXII\ 7611\,\text{\AA} has been identified in just one extragalactic object. We, therefore, increased by a factor 3 the number of sources where such a line is detected.

\subsection{Physical conditions of the NLR in CLiF AGNs}

\begin{table*}
\centering
\caption{Density and temperature values for the NLR determined from sensitive emission-line flux ratios 
}
\begin{tabular}{lcccccc}
\hline
\multicolumn{1}{c}{Galaxy} & Density $_{ [\mathrm{S \ II]}}$ & Density $_{ [\mathrm{Ar \ IV}]}$ & Temperature $_{ [\mathrm{O \ III}]}$ & Temperature $_{ [\mathrm{N \ II]}}$ & Temperature $_{ [\mathrm{S \ II}]}$ & Temperature $_{ [\mathrm{S \ III}]}$ \\ 
 & (cm$^{-3}$) & (cm$^{-3}$) & (K) & (K) & (K) & (K)\\
\hline
ESO 138 G1                 & 449                & 3382                & 16396                   & 7490 - 10074           & 4350 - 6133            & does not converge       \\
SDSS J164+43   & -                  & -                   & 63021                   & -                      & -                      & -                       \\
III zw 77                  & 384                & -                   & 33620                   & 11539 - 18097          & 4598 - 10520           & 19221 - 20752           \\
MRK 1388                   & 964                & 16753               & -                       & 13187 - 22055          & 6214                   & 13187 - 22055           \\
SDSS J124+44   & 652                & -                   & does not converge       & -                      & -                      & -                       \\
NGC 424                    & 542                & 2588                & 17499                   & 7773 - 10563           & 3679 - 6215            & -                       \\ \hline
\end{tabular}
\label{tab:condfi}
\end{table*}

In order to estimate the physical conditions of the NLR gas (electron density, $n_{\rm e}$, and temperature, $T_{\rm e}$) several diagnostic line ratios were employed. Previously, all measured fluxes for the NLR were corrected for reddening, as derived in the previous section. The gas density was evaluated using two diagnostics: the \SII\ 6716/6730 \text{\AA} and the \ArIV\ 4711/4740 \text{\AA} emission-line flux ratios. The former is suitable for gas density below 10$^4$ cm$^{-3}$ as the line ratio saturates when $n_{\rm e}$ is above that value. The later is suitable for mapping gas of higher density ($< 10^5$ cm$^{-3}$) because the lines involved have larger critical densities and IP than those of [\ion{S}{ii}]. Note that both diagnostics are insensitive to electron temperatures within a very large interval. In this calculation, we followed the procedure and relationships outlined in \cite{proxauf/2014}, assuming always a temperature of 10$^4$~K. Varying the temperature by a factor of 3 up and down has very little effect on the results. 

Electron temperatures were determined using single-ion emission pairs of lines that have excitation levels with a considerable difference in energy. A good example is the emission-line flux ratio \OIII\ (4959+5007)/4363 \text{\AA}.  

Other temperature sensitive line ratios are \NII\ (6548+6583)/5754 \text{\AA}, \SII\ (6716+6730)/4068 \text{\AA} and \SIII\ (0.906+0.953) $\mu$m/6312 \text{\AA}. The simultaneous detection of all these lines in the spectra of CLiF AGN allowed us to compare the physical conditions resulting from different diagnostics. For \OIII\, the expressions derived by \cite{proxauf/2014} assuming an electron density of 10$^{4}$ cm$^{-3}$ were employed.  The gas temperature from the three latter ratios we determined using the TEMDEN task included in the STSDAS version 3.18.3 package of IRAF. The results obtained from the electron density and temperature diagnostics are shown in Table \ref{tab:condfi}.

Our results evidence $n_{\rm e}$ in the interval 10$^{2}$-10$^{4}$ cm$^{-3}$, in agreement to values found by other authors (\citealt{bennert/2006}; \citealt{kakkad/2018}) in samples with non-CLiF AGN. It implies that the bulk of the NLR in CLiF AGN is dominated by gas with density very similar to that found in other AGNs. The enhancement of coronal lines seen in these objects must be due to other physical properties.

Regarding the temperature, it can be verified that all values derived from \OIII\ suggest $T_{\rm e} \sim$ 1.5$-$6$ \times 10^{4}$~K. The other three sensitive temperature diagnostics show values in the interval ($1-2$) $\times 10^{4}$~K.
This result further supports our hypothesis of a matter-bounded dominated NLR. According to \citet{binette/1997}, the temperature in the MB clouds, where most of the \OIII\ and CLs would be produced, exceeds at least by 5000~K the temperature of IB clouds, where the bulk of [\ion{N}{ii}], [\ion{S}{ii}] and other low-ionization species are produced. Moreover, according to \cite{yan/2018}, temperatures around 10$^{4}$~K are associated with the gas photoionized by the central source while values as large as 10$^{5}$ K are typically associated with the shocked gas. No evidence of such high $T_{\rm e}$ were found here. Thus, our results suggest that the NLR gas of CLiF AGN is primarily photoionized by the AGN although the presence of shocks is not discarded. Similar results were found by \citet{rodriguez/2011} in the study of the CLR in a sample of non-CLiF AGN. 

We should note at this point that the physical conditions derived for the NLR can be extended to the CLR under the assumption that both regions share the same kinematics, chemical composition, physical extension and are illuminated by the same ionizing continuum. Otherwise, our results here do not necessarily reflect the conditions of the gas producing the CLs. The lack of sensitive diagnostics of gas temperature involving high-ionization lines hinders the correct characterization of the CLR.  In this regard, the observation of CLiF AGN using adaptive optics (AO) and higher spectral resolution, at least for the closest sources, is fundamental to set firm constrains on the size and physical properties of the region emitting the CLs in these objects. Non-CLiF AGN observed at spatial scales of $\sim 0.15"$ with AO systems and/or space-base observatories (i.e The Hubble Space Telescope) show CL emission co-spatial with [\ion{O}{iii}], [\ion{S}{ii}] and \ion{H}{i} gas in the inner portions of the NLR \citep{mazzalay/2010, muller/2011, rodriguez/2017, may/2018}. These results warrant that, on a first approximation, the values of electron density and temperature derived for the NLR can be extended the CLR.

\section{kinematics}\label{kinematics}

\subsection{Gas kinematics}

Earlier studies using integrated long-slit spectroscopy found out that CLs tend to be broader than forbidden low-ionization lines and their centroid position blueshifted relative to the systemic velocity of the galaxy (\citealt{wilson/1979}; \citealt{pelat/1981}; \citealt{evans/1988}). Furthermore, the presence of correlations between the width of the lines and the corresponding ionization potential and/or critical density of the transitions were taken as evidence of a stratification in the ionization state of the gas, in the sense that the high-ionization gas would be located in the inner portions of the NLR while  low-ionization gas farther out. The discovery of blue asymmetries, split line profiles, and gas kinematics largely departing from rotation in the CLs suggested that at least part of the coronal gas is associated with outflowing material from the NLR (\citealt{erkens/1997};  \citealt{rodriguez/2002}; \citealt{rodriguez/2006}; \citealt{muller/2011}; \citealt{may/2018}). 

However, not all AGN with CLs display such extreme properties. \citet{rodriguez/2011} found that the positive correlation between FWHM and IP was detected in 10 out of 54 galaxies examined. In some objects of their sample, no dependence at all between these two quantities was found while in some others, CLs have widths comparable to or even smaller than those displayed by low-ionization lines. Similar results were previously reported by \cite{knop/1996}.  

In CLiF AGN, \citetalias{rose/2015a} did not identify a correlation between FWHM and IP in none of the galaxies examined. In all cases, the distribution of FWHM values tend to be rather similar for low and high ionization potential lines. Due to the availability of lines of higher ionization potential in the NIR and the larger spectral resolution of our data compared to those of SDSS, we revisited this issue. 

\begin{figure*}

	\begin{center}
    \includegraphics[width=\textwidth]{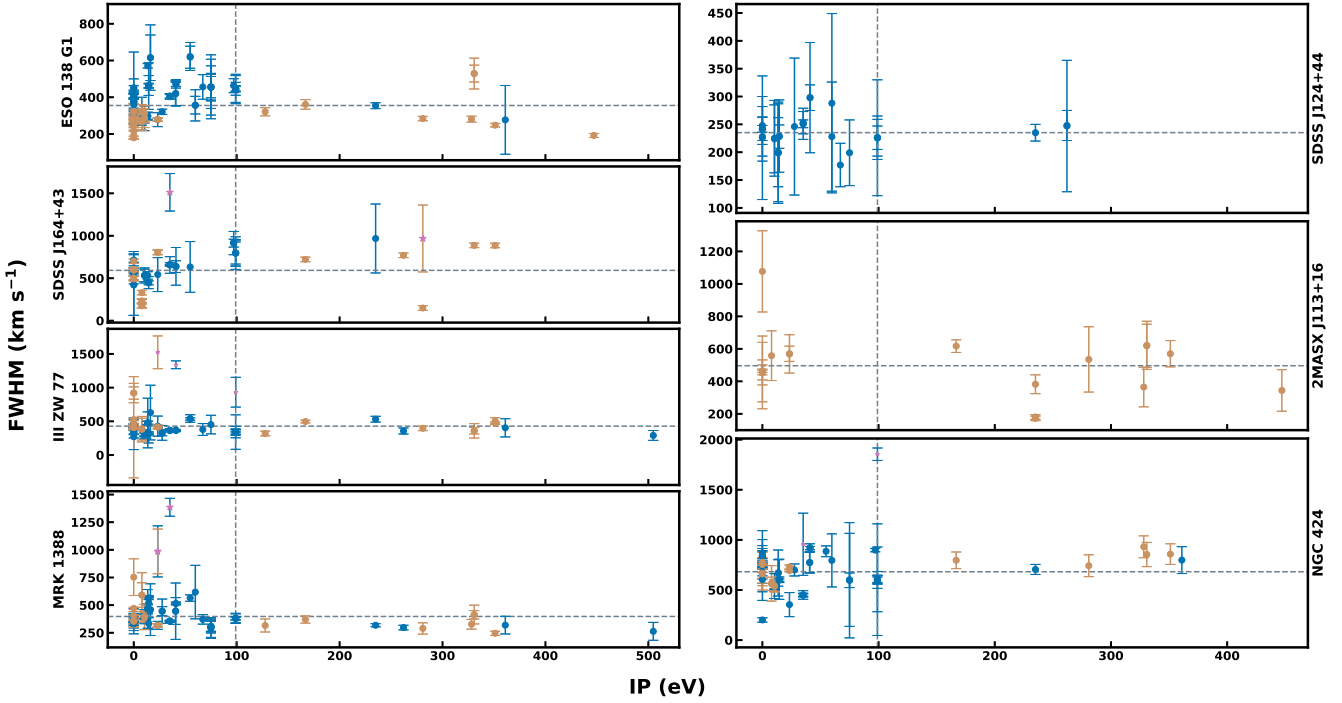}
    \caption{FWHM vs IP of the emission lines identified in the sample of CLiF AGN. The blue dots represent the measurements for optical lines while brown dots are for NIR lines. The vertical dashed line represents the boundary for a line to be considered as coronal (IP $\geq 100$~eV). The horizontal dashed line is average value of the FWHM for each source.}\label{fig:cinematica}
    \end{center}
	
	\centering

\end{figure*}

Fig. \ref{fig:cinematica} plots the FWHM vs IP for the 7 CLiF AGNs of the sample. Even with the addition of new lines of high IP,  there is no clear distinction between the FWHM of coronal lines and that of the low ionization potential lines, confirming previous results of \citetalias{rose/2015a}. Indeed, the distribution of values of FWHM found in CLiF AGN are similar to that measured by \citet{rodriguez/2011} for non-CLiFs.   Fig.~\ref{fig:cinematica} shows that NIR CLs display FWHM from 250~km\,s$^{-1}$ (ESO\,138\,G1 and Mrk\,1388) up to values larger than 1000~km\,s$^{-1}$ (SDSS\,J164+43 and NGC\,424). This same range in line width is also observed in low-ionization lines. This suggests that the velocity field of the NLR in CLiF and non-CLiF AGN may be intrinsically the same.

Our data also confirm the lack of any trend between line shifts from the systemic velocity, $\Delta$V, and the IP   reported by \citetalias{rose/2015a}. Column~3 of Tables~\ref{tab:esoopt_tabelao} to~\ref{tab:ngcinfra_tabelao} lists $\Delta$V (in km\,s$^{-1}$) for each emission-line detected in the sample. It can be seen that the measured shift of the forbidden and permitted lines fluctuates regardless of the IP of the line. No trend is also detected between $\Delta$V and FWHM.   

Our results suggest that the region where the coronal line forest is produced shares the same kinematics as the gas responsible for the emission of low-ionization lines. 
Under the assumption of a virialized velocity field and a stratification in  the ionization state of the gas, in the sense that high-ionization lines would be produced closer to the central source (i.e the torus) and lower-ionization lines outwards, we would expect a positive correlation between FWHM and IP, which is not observed. The lack of such correlation, however, does not make CLiF AGN special as similar results were previously related in many other non-CLiF AGN \citep{knop/1996, rodriguez/2011}.

We highlight that in none of the sources examined here, extended emission from the NLR was detected. In all cases, the NLR  was restricted to sizes $<$2" in the spatial direction of the slit. For the closest objects (ESO~138-G~001, NGC\,424, Mrk\,1388 and III\,Zw\,77), it translates to scales down to $\sim$400~pc. Therefore, the NLR of these objects should be very compact. If no stratification in gas ionization exists and the velocity field is not virialized, under seeing-limiting conditions no correlation between FWHM and IP is expected.   In order to confirm this scenario, observations of CLiF AGN with telescopes equipped with adaptic optics systems that allow to resolve angular scales down to tens of parsecs and at spectral resolutions of R$>$4000  are necessary. These data would allow to put firm constrains to the kinematics and likely location of the CLR in such objects.

\subsection{Stellar velocity dispersion and black hole mass determination} \label{sec:smbh}

 \begin{figure}

    \begin{center}
    \includegraphics[width=\columnwidth,height=7cm]{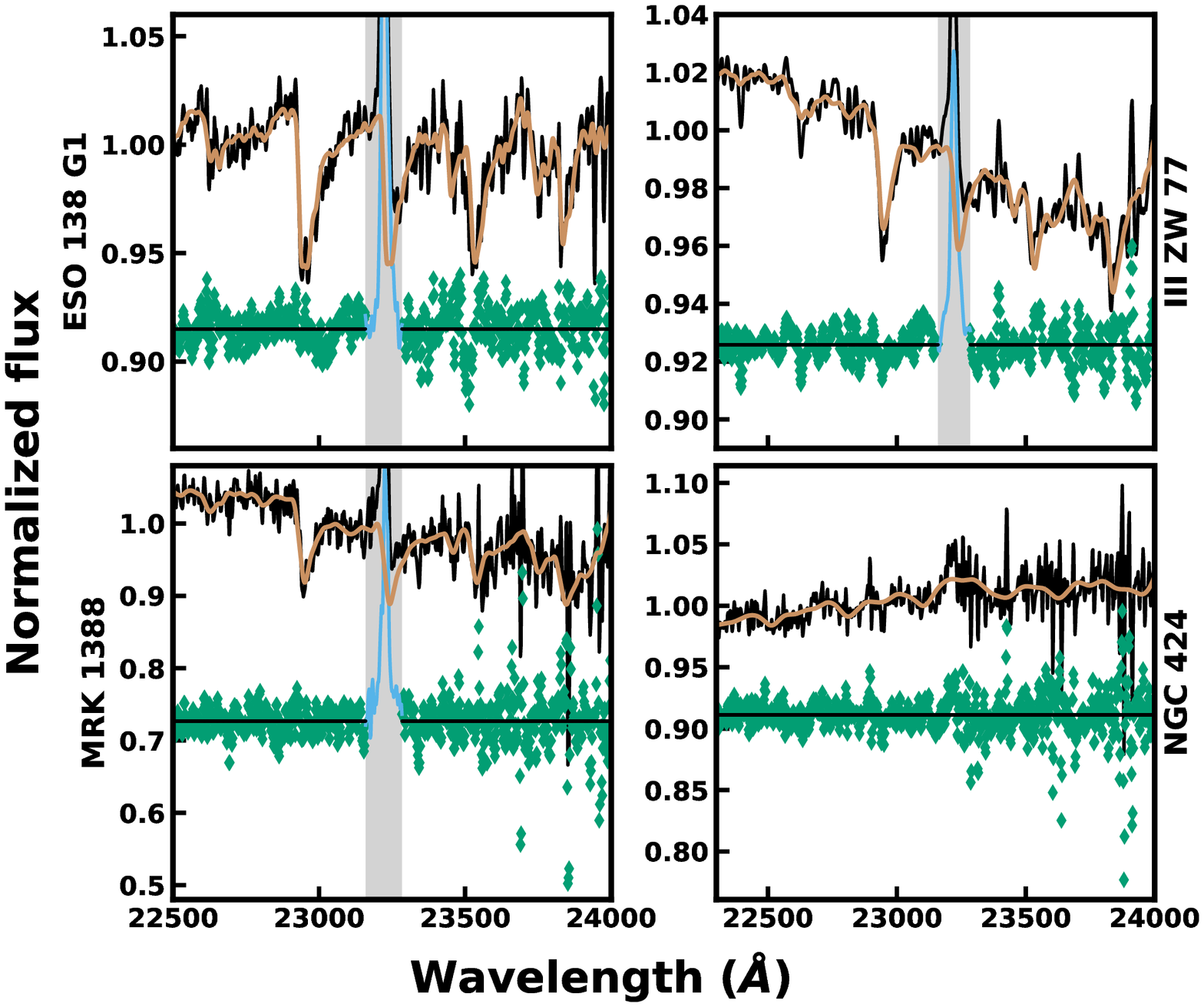}
    
    \caption{Fit of pPXF to the region around the 2.32~$\mu$m CO absorptions. The black line is the observed spectrum while the brown line is the best fit. Regions masked from the fit are in blue and residuals are in green.}\label{fig: CO}
    
    \end{center}

	\centering
\end{figure}

\begin{figure}

    \begin{center}
    \includegraphics[width=\columnwidth,height=7cm]{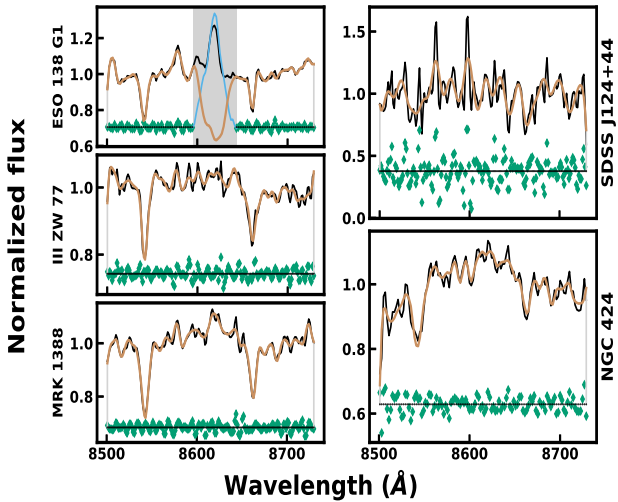}
    \caption[Ajuste do pPXF na região das absorções do CaT]{The same as Fig. \ref{fig: CO} for the region around the CaT absorptions}\label{fig: CaT}
    \end{center}
	
	\centering
\end{figure}

\begin{table*}
 \centering
  \caption{Values of $\sigma$ in units of km s$^{-1}$ and the  values obtained for the masses of the SMBHs (M$_{\odot}$) by means of the M-$\sigma$  ratio \citep{kormendy/2013}.} \label{tab: smbh}
\begin{threeparttable}
 
\begin{tabular}{lcccc}
\hline
\multicolumn{1}{c}{Galaxy} & $\sigma_{CO}$     & $\sigma_{CaT}$   & $\sigma$ & M$_{\textrm{SMBH}}$ \\ \hline
ESO 138 G1                 & 124.90$\pm$7.01   & 131.03$\pm$14.30 & 127.96$\pm$7.96  & 4.37$_{-1.02}^{+1.14}$ $\times 10^{7}$ \\[2pt]
SDSS J164+43               & -                 & -                & -                & 2.7$\pm$0.4 $\times 10^{7}$*        \\[2pt]
III ZW 77                  & 166.99$\pm$14.72  & 161.59$\pm$10.17 & 164.29$\pm$12.44 & 1.31$_{-0.10}^{+0.11}$ $\times 10^{8}$ \\[2pt]
MRK 1388                   & 160.41$\pm$20.55** & 112.95$\pm$8.30  & 112.95$\pm$8.30  & 2.53$_{-1.02}^{+1.14}$ $\times 10^{7}$ \\[2pt]
SDSS J124+44               & -                 & 119.97$\pm$18.43 & 119.97$\pm$18.43 & 3.29$_{-1.02}^{+1.15}$ $\times 10^{7}$ \\[2pt]
NGC 424                    & -                 & 146.65$\pm$12.76 & 146.65$\pm$12.76 & 7.94$_{-1.02}^{+1.14}$ $\times 10^{7}$ \\ \hline
\end{tabular}
 \begin{tablenotes}
      \small
      \item * For this object, it is not possible to identify stellar absorption signatures, the power law that describes the AGN emission has a greater influence on the continuum (see figures \ref{fig: ppxfFig_opt} and \ref{fig: ppxfFig}). Thus we employed the virial relation of \cite{vestergaard/2006}, which relates the black hole mass to the brightness of the AGN emission continuum at 5100 \text{\AA} and the width of the BLR hydrogen lines.
      \item ** As can be seen in Fig. \ref{fig: CO}, the K-band spectrum does not have well-defined CO absorptions, so this value was not used in the analysis.
    \end{tablenotes}
  \end{threeparttable}
\end{table*}

Part of the spectroscopic properties displayed by CLiF AGN can be explained if the mass of the central black hole is small ($M_{\rm BH} < 10^7$ M$\odot$). Assuming that the CLR is photoionized by the continuum emitted primarily by the accretion disk, AGNs with smaller mass black holes will have hotter disks. In such a case, the peak of the spectral energy distribution is shifted towards higher energies. Under this scenario, coronal lines will be favoured because of the large supply of photons with energies in the X-ray domain. This idea was explored by \cite{cann/2018}, who propose that intermediate mass black hole galaxies ($10^3 < M_{\rm BH} < 5 \times 10^6$) M$\odot$ can be hunted by the detection of bright coronal lines in the NIR and MIR region.  

In this work, we determine $M_{\rm SMBH}$ from the stellar velocity dispersion $- \sigma_*$ obtained from the stellar population fit in the optical and NIR by means of the pPXF. The relationship proposed by \cite{kormendy/2013} was employed.

In order to measure $\sigma_*$ in the NIR, we employed the Gemini NIR late-type stellar library (\citealt{winge/2009}), which contains 60 individual stellar spectra with spectral types ranging from F7III to M5III, observed in the K band with resolution  of $\sim$ 3,2 \text{\AA} (FWHM). For the optical spectra, the region containing the \ion{Ca}{ii} triplet was employed. In this case, we used the stellar templates of \cite{cenarro/2001}, which contains 706 stellar spectra with 1.5 \text{\AA} (FWHM) spectral resolution. These spectra are part of the \textit{Medium-resolution Isaac Newton Telescope library} (MILES) (\citealt{sanchez/2006}).

The fits carried out with pPXF for the region of the CO and CaT absorption are shown in Figures \ref{fig: CO} and \ref{fig: CaT}, respectively. Table \ref{tab: smbh} displays $\sigma$ obtained and the corresponding mass of the SMBH ($M_{\rm SMBH}$) for each galaxy. When two measurements of $\sigma$ were available, we took the average of both values and the result was employed to determine $M_{\rm SMBH}$. From our sample of 7 galaxies, we were able to measure the mass of the central black hole in 6 of them. Note that we do not employed the virial relationship because the strength of the broad component in H$\beta$ or even in H$\alpha$ was weak and that could introduce a bias in the $M_{\rm SMBH}$ determination.

Table~\ref{tab: smbh} shows that CLiF AGN display values of $M_{\rm SMBH}$ over a very narrow interval of $M_{\rm SMBH}$, between 10$^7$ and 10$^8$~M$\odot$. NGC\,3786, NGC\,4151, and NGC\,5548 are among bona-fide Type~I AGN with $M_{\rm SMBH}$ in the above range (\citealt{bentz/2015}). They display a very rich CL spectrum, both in the optical and the NIR \citep{evans/1988, knop/1996, rodriguez/2011}. Therefore, the CLiF nature of our sources cannot be ascribed to a mere effect of the central black-hole mass. It is certainly worth to explore other properties such as bolometric luminosity, Eddington ratio, and accretion rate for CLiF and non-CLiF sources in order to detect any trend that could explain the peculiar properties of the former. We left this analysis for a future work and it is first mandatory to expand the number of sources with CLiF properties for a meaningful statistics.

\section{Final remarks}\label{conclusions}

In the contemporary scenario, studying the forbidden high-ionization lines in AGNs is of great importance in order to gather crucial insights about the most energetic processes at the inner portion of the NLR. Here, we carried out a multi-wavelength analysis of a selected group of AGNs known to display a prominent coronal line forest spectrum. In addition to optical data, we include the first report of NIR spectroscopy on these objects in the literature. The combination of optical and NIR spectroscopy aims at understanding the physical processes at work in the central parsecs of these objects and unveiling if CLiF AGN are indeed a special class of active galactic nuclei. The main results gathered from this work are summarised below.

\begin{itemize}

\item{The optical and NIR continuum emission in CLiF AGN observed under seeing-limited conditions}   is  strongly affected by the underlying stellar population of the host galaxy. Thus, in order to study the gas emission, it is first necessary to model and subtract that component. In the NIR, in addition to the stellar contribution, a continuum emission due to dust heated at temperatures between 700 and 1500~K by the AGN   is also present. It dominates the $H$-band continuum and redwards. 

\item{The analysis of the emission-line spectrum showed the presence of broad components with FWHM $\geq$2000~km\,s$^{-1}$ in the permitted lines in 3 out of 4 galaxies previously classified as Type~II AGN. This result points out that the sample of CLiF AGN is dominated by Type~I objects instead of Type~II.}

\item{Due to the increase of wavelength coverage by including NIR spectra, we employed 5 different diagnostics of extinction to assess if dust is present in the NLR. We found that CLiF AGN are affected by reddening in the nuclear region, with values ranging from 0.12 $\leq$ E(B-V) $\leq$ 0.57. In this determination, the subtraction of the stellar population proved to be critical, as H$\beta$ may be strongly diluted by the presence of that component.}

\item{We derived the emission-line flux ratios [\ion{Fe}{xi}]$\lambda$7890/[\ion{Fe}{x}]$\lambda$6374 and [\ion{Fe}{x}]$\lambda$6374/[\ion{Fe}{vii}]$\lambda$6087 in the optical and [\ion{Si}{vi}]~1.963$\mu$m/[\ion{Si}{x}]~1.43$\mu$m and [\ion{S}{viii}]~0.991$\mu$m/[\ion{S}{ix}]~1.251$\mu$m in the NIR for the CliF AGN. They were compared to those measured in normal AGN with CL emission. The analysis showed that the two groups of AGN are not distinguished  regarding the relative strength of these coronal lines. However, it is interesting to note that CLiF AGN display simultaneously all coronal lines previously identified in other AGNs. This strongly contrast to what is found in samples of non-CLiF AGN, where less than 5\% of the galaxies display lines such as [\ion{Fe}{xiii}] or [\ion{S}{xi}]. Moreover, CLiF AGN display a conspicuous forest of [\ion{Fe}{v}], [\ion{Fe}{vi}], and [\ion{Fe}{vii}] lines in the optical region, rarely seen in classical AGN.}

\item{The remarkable CL spectrum displayed by CLiF AGN and their strong intensity relative to low-ionization lines suggest that these sources have a NLR dominated by matter-bounded clouds. Such clouds are responsible for the emission of high-excitation lines. Ionization-bounded clouds, where the low-ionization lines are emitted, should be outnumbered by the former.}

\item{Temperature and density diagnostics were employed to characterise the physical state of the NLR in the CLiF AGN sample. We found values of $T_{\rm e}$ in the range 10$^{4} - 6.3\times 10^{4}$~K, measured from the [\ion{O}{iii}] and [\ion{S}{iii}] lines.  Temperature diagnostics using low-ionization lines ([\ion{N}{ii}] and [\ion{S}{ii}]) suggest smaller $T_{\rm e}$ of $\sim 10^{4}$~K. This result further supports our hypothesis  of a matter-bounded dominated NLR. Values of gas density ($n_{\rm e}$) using Sulfur and Argon lines indicate values close 500~cm$^{-3}$ and up to 1.6$\times 10^{4}$~cm$^{-3}$, respectively. Overall,  the NLR ionization mechanism is associated with photoionization by the central source although the presence of shocks is not fully discarded.}

\item{The kinematics of the NLR show that the coronal line gas shares similar  characteristics to that emitting low ionization lines. For instance, no trend between FWHM and IP is observed, even after considering  several lines with IP $>$100~eV and up to $\sim$500~eV. Thus, it is not possible to asses if the gas responsible for the coronal emission is located in the inner portion of the NLR. However, the lack of extended NLR emission in all CLiF suggests that the bulk of the NLR gas is nuclear, and subjected to our angular resolution, presumably within the inner 500~pc.
}

\item{We determined the mass of the supermassive black hole $-$SMBH, using the M-$\sigma$ relationship and the virial scaling using the FWHM of H$\beta$ and the continuum luminosity at 5100~\AA. All masses were found to be in a very narrow range of values, between 10$^{7}$ - 10$^{8}$ M$_{\odot}$. This result rules out the hypothesis that the CLiF is due to a very hot accretion disk, i.e., that produced by a small black hole mass AGN.}

\end{itemize}

We conclude that CLiF AGN are not a separate class of AGN due to the enhancement of the forbidden high-ionization line spectrum. However, when compared to non-CLiF AGN, they usually occupy the high-end of the distribution either in CL luminosity or when a line flux ratio between a given CL and \ion{H}{i} lines is considered.  It is fundamental to identify a larger number of sources with similar characteristics to elaborate more definitive conclusions about the properties of the NLR in these objects. Moreover, the observation of the existing sources, at least the closest ones, at larger angular resolutions will strongly help to put firm constrains to the size of the NLR and of the coronal line region. 

\section*{Acknowledgements}

We thank to the annonymous referee for his/her useful comments and suggestions to improve this manuscript. FCCC acknowledges the PhD grant from CAPES. RR thanks CNPq, CAPES and FAPERGS for partially funding this project. ARA acknowledges CNPq for partial support to this project. Based on observations obtained at the Gemini Observatory, which is operated by the Association of Universities for Research in Astronomy, Inc., under a cooperative agreement with the NSF on behalf of the Gemini partnership: the National Science Foundation (United States), National Research Council (Canada), CONICYT (Chile), Ministerio de Ciencia, Tecnolog\'{i}a e Innovaci\'{o}n Productiva (Argentina), Minist\'{e}rio da Ci\^{e}ncia, Tecnologia e Inova\c{c}\~{o}es (Brazil), and Korea Astronomy and Space Science Institute (Republic of Korea). This paper is also based on observations obtained at the Southern Astrophysical Research (SOAR) telescope, which is a joint project of the Minist\'{e}rio da Ci\^{e}ncia, Tecnologia, Inova\c{c}\~{o}es e Comunica\c{c}\~{o}es (MCTIC) do Brasil, the U.S. National Optical Astronomy Observatory (NOAO), the University of North Carolina at Chapel Hill (UNC), and Michigan State University (MSU).

\section*{Data availability}

The data underlying this article will be shared on reasonable request to the corresponding author.

%%%%%%%%%%%%%%%%%%%%%%%%%%%%%%%%%%%%%%%%%%%%%%%%%%



\bibliographystyle{mnras}
\bibliography{bib} % if your bibtex file is called example.bib
%\usepackage[style=authoryear-comp, maxnames=2, uniquelist=false]{biblatex}

%\begin{thebibliography}{99}
%\end{thebibliography}

%%%%%%%%%%%%%%%%%%%%%%%%%%%%%%%%%%%%%%%%%%%%%%%%%%

%%%%%%%%%%%%%%%%% APPENDICES %%%%%%%%%%%%%%%%%%%%%

\appendix

\section{OPTICAL EMISSION-LINE TABLES}

\begin{table*}

   \centering
  
   \caption{Emission-line flux to the galaxy ESO 138-G001 in the optical. FWHM is in units of km s$^{-1}$ and the flux in $10^{-14}$  erg s$^{-1}$ cm$^{-2}$}
\begin{tabular}{lcccccccccc}
\cmidrule{1-7}     \multicolumn{1}{c}{Ion} & $\lambda_{obs}$ (\text{\AA}) & $\Delta v$ (km s$^{-1}$) & IP (eV) & FWHM & Flux  & Corrected flux &   &   &   &  \\
\cmidrule{1-7}     $[$Ne V$]$   & 3424.8 & -94.6  & 97.1  & 463$\pm$38  & 19.48$\pm$1.60  & 192.05$\pm$15.80 \\
$[$O II$]$   & 3726.5 & 37.7   & 13.6  & 571$\pm$12  & 71.82$\pm$0.79  & 637.91$\pm$7.05  \\
$[$Fe VII$]$ & 3758.9 & -1.6   & 99.0  & 445$\pm$80  & 5.00$\pm$0.90   & 43.94$\pm$7.87   \\
$[$Ne III$]$ & 3868.0 & -58.9  & 41.0  & 471$\pm$12  & 37.41$\pm$0.72  & 315.76$\pm$6.05  \\
$[$Fe V$]$   & 3888.8 & -191.1 & 54.8  & 618$\pm$60  & 7.73$\pm$0.75   & 64.73$\pm$6.25   \\
$[$Ne III$]$ & 3967.7 & 17.4   & 41.0  & 473$\pm$23  & 15.30$\pm$0.74  & 124.00$\pm$5.99  \\
$[$S II$]$   & 4067.5 & -81.1  & 10.4  & 285$\pm$25  & 3.47$\pm$0.31   & 26.89$\pm$2.37   \\
\end{tabular}%
     
   \label{tab:esoopt_tabelao}%
 \end{table*}%

\begin{table*}
   \centering
   \caption{Emission-line flux to the galaxy SDSS J164126.90+432121.5 in the optical. FWHM is in units of km s$^{-1}$ and the flux in $10^{-15}$  erg s$^{-1}$ cm$^{-2}$}
     \begin{tabular}{lcccccccccc}
\cmidrule{1-7}     \multicolumn{1}{c}{Ion} & $\lambda_{obs}$ (\text{\AA}) & $\Delta v$ (km s$^{-1}$) & IP (eV) & FWHM & Flux & Corrected flux &   &   &   &  \\
\cmidrule{1-7}     $[$Ne V$]$   & 3343.6 & -199.0 & 97.1  & 914$\pm$137  & 1.40$\pm$0.21 & 4.01$\pm$0.60  \\
$[$Ne V$]$   & 3423.7 & -190.9 & 97.1  & 913$\pm$46   & 4.16$\pm$0.21 & 11.71$\pm$0.60 \\
$[$O II$]$   & 3727.1 & 85.9   & 13.6  & 533$\pm$44   & 1.08$\pm$0.09 & 2.90$\pm$0.24  \\
$[$Fe VII$]$ & 3757.0 & -153.1 & 99.0  & 797$\pm$152  & 0.68$\pm$0.13 & 1.81$\pm$0.34  \\
$[$Ne III$]$ & 3868.6 & -12.4  & 41.0  & 637$\pm$69   & 1.58$\pm$0.17 & 4.15$\pm$0.44  \\
$[$Fe V$]$   & 3890.2 & -83.2  & 54.8  & 633$\pm$299  & 0.36$\pm$0.17 & 0.93$\pm$0.44  \\
$[$Ne III$]$ & 3968.5 & 77.8   & 41.0  & 640$\pm$222  & 0.49$\pm$0.17 & 1.26$\pm$0.45  \\
\end{tabular}%
   \label{tab:j164opt_tabelao}%
 \end{table*}%

\begin{table*}
   \centering
   \caption{Emission-line flux to the galaxy III ZW 77 in the optical. FWHM is in units of km s$^{-1}$ and the flux in $10^{-15}$  erg s$^{-1}$ cm$^{-2}$}
     \begin{tabular}{lcccccccccc}
\cmidrule{1-7}     \multicolumn{1}{c}{Ion} & $\lambda_{obs}$ (\text{\AA})& $\Delta v$ (km s$^{-1}$) & IP (eV) & FWHM & Flux & Corrected flux &   &   &   &  \\
\cmidrule{1-7}     $[$O II$]$ & 3726.1 & 5.5     & 13.6  & 476$\pm$24   & 13.22$\pm$0.66  & 69.80$\pm$3.48   \\
$[$Fe VII$]$*                                  & 3757.0 & -153.1  & 99.0  & 931$\pm$411  & 5.21$\pm$2.30   & 27.33$\pm$12.07  \\
$[$Fe VII$]$                                   & 3758.6 & -25.5   & 99.0  & 341$\pm$48   & 5.91$\pm$0.84   & 30.99$\pm$4.42   \\
$[$Fe V$]$                                     & 3836.3 & -231.9  & 54.8  & 542$\pm$56   & 2.71$\pm$0.28   & 13.91$\pm$1.46   \\
$[$Ne III$]$                                   & 3867.9 & -66.6   & 41.0  & 372$\pm$6    & 26.60$\pm$0.20  & 135.28$\pm$1.00  \\
$[$Fe V$]$                                     & 3890.3 & -75.5   & 54.8  & 535$\pm$18   & 8.56$\pm$0.28   & 43.24$\pm$1.44   \\
$[$Ne III$]$                                   & 3967.1 & -28.0   & 41.0  & 363$\pm$16   & 4.50$\pm$0.20   & 22.19$\pm$0.97   \\
\end{tabular}%
   \label{tab:iiizw77opt_tabelao}%
 \end{table*}%

\begin{table*}
   \centering
    \caption{Emission-line flux to the galaxy MRK 1388 in the optical. FWHM is in units of km s$^{-1}$ and the flux in $10^{-15}$  erg s$^{-1}$ cm$^{-2}$}
     \begin{tabular}{lcccccccccc}
\cmidrule{1-7}     \multicolumn{1}{c}{Ion} & $\lambda_{obs}$ (\text{\AA}) & $\Delta v$ (km s$^{-1}$) & IP (eV) & FWHM & Flux & Corrected flux &   &   &   &  \\
\cmidrule{1-7}     $[$Ne III$]$ & 3868.4 & -27.9  & 41    & 513$\pm$6    & 51.75 $\pm$ 0.52  & 86.61 $\pm$ 0.87  \\
$[$Fe V$]$                                       & 3889.6 & -129.4 & 54.8  & 564$\pm$29   & 11.09 $\pm$ 0.58  & 18.52 $\pm$ 0.97  \\
$[$Ne III$]$                                     & 3967.9 & 32.5   & 41    & 515$\pm$13   & 21.83 $\pm$ 0.54  & 36.18 $\pm$ 0.89  \\
$[$S II$]$                                       & 4069.0 & 29.5   & 10.4  & 397$\pm$24   & 8.67 $\pm$ 0.52   & 14.21 $\pm$ 0.85  \\
$[$S II$]$                                       & 4076.7 & 25.8   & 10.4  & 396$\pm$64   & 3.21 $\pm$ 0.52   & 5.26 $\pm$ 0.85   \\
H$\delta$*                                       & 4099.8 & -141.9 & 0     & 1024$\pm$171 & 6.15 $\pm$ 1.03   & 10.05 $\pm$ 1.68  \\
H$\delta$                                        & 4101.7 & -3.1   & 0     & 336$\pm$19   & 5.9 $\pm$ 0.34    & 9.63 $\pm$ 0.56   \\
\end{tabular}%
   \label{tab:mrkopt_tabelao}%
 \end{table*}%

 \begin{table*}
   \centering
   \caption{Emission-line flux to the galaxy J124134.25+442639.2 in the optical. FWHM is in units of km s$^{-1}$ and the flux in $10^{-16}$  erg s$^{-1}$ cm$^{-2}$}
     \begin{tabular}{lcccccccccc}
\cmidrule{1-7}     \multicolumn{1}{c}{Ion} & $\lambda_{obs}$ (\text{\AA}) & $\Delta v$ (km s$^{-1}$) & IP & FWHM & Flux & Corrected flux &   &   &   &  \\
\cmidrule{1-7}     $[$O II$]$ & 3729.2 & -54.3  & 13.6  & 199$\pm$88  & 2.62$\pm$1.16  & 8.43$\pm$3.73  \\
$[$Fe VII$]$                                   & 3759.0 & 6.4    & 99.0  & 226$\pm$39  & 7.02$\pm$1.20  & 22.46$\pm$3.84 \\
$[$Ne III$]$                                   & 3868.9 & 10.8   & 41.0  & 298$\pm$23  & 11.26$\pm$0.88 & 35.39$\pm$2.77 \\
$[$Ne III$]$                                   & 3967.6 & 9.8    & 41.0  & 298$\pm$99  & 2.70$\pm$0.90  & 8.27$\pm$2.76  \\
H$\gamma$                                      & 4340.1 & -25.6  & 0.0   & 242$\pm$58  & 3.35$\pm$0.80  & 9.32$\pm$2.22  \\
$[$O III$]$                                    & 4363.0 & -14.4  & 35.1  & 251$\pm$28  & 6.73$\pm$0.75  & 18.59$\pm$2.07 \\
He II                                          & 4685.8 & 5.8    & 0.0   & 248$\pm$34  & 5.16$\pm$0.71  & 13.05$\pm$1.79 \\
\end{tabular}%
   \label{tab:tabelaoJ124}%
 \end{table*}%

\begin{table*}
   \centering
   \caption{Emission-line flux to the galaxy NGC 424 in the optical. FWHM is in units of km s$^{-1}$ and the flux in $10^{-14}$  erg s$^{-1}$ cm$^{-2}$}
    \resizebox{!}{11.5cm}{\begin{tabular}{lcccccccccc}
\cmidrule{1-7}     Ion & $\lambda_{obs}$ (\text{\AA}) & $\Delta v$ (km s$^{-1}$) & IP (eV)  & FWHM & Flux  & Corrected flux &   &   &   &  \\
\cmidrule{1-7}     $[$Ne V$]$ & 3419.3 & -122.9 & 97.1 & 898$\pm$17   & 29.32$\pm$0.55  & 84.59$\pm$1.59  \\
$[$Fe VII$]$*                                  & 3575.2 & -562.8 & 99   & 1855$\pm$250 & 1.70$\pm$1.18   & 4.78$\pm$3.32   \\
$[$Fe VII$]$                                   & 3581.4 & -44.6  & 99   & 600$\pm$84   & 2.72$\pm$0.38   & 7.64$\pm$1.07   \\
H I                                            & 3722.6 & 384.8  & 0    & 743$\pm$12   & 21.67$\pm$0.35  & 59.58$\pm$0.96  \\
$[$Fe VII$]$*                                  & 3748.3 & -434.2 & 99   & 1862$\pm$629 & 3.73$\pm$1.26   & 10.21$\pm$3.45  \\
$[$Fe VII$]$                                   & 3754.5 & 60.3   & 99   & 603$\pm$45   & 5.44$\pm$0.41   & 14.88$\pm$1.12  \\
H I                                            & 3792.4 & -110.5 & 0    & 744$\pm$348  & 0.77$\pm$0.36   & 2.09$\pm$0.98   \\
\end{tabular}}%
   \label{tab:ngcopt_tabelao}%
 \end{table*}%

\section{NEAR INFRARED EMISSION-LINE TABLES}

\begin{table*} 
   \centering
   \caption{Emission-line flux to the galaxy ESO 138 G1 in the NIR. FWHM is in units of km s$^{-1}$ and the flux in $10^{-15}$  erg s$^{-1}$ cm$^{-2}$}
     \begin{tabular}{lcccccccccc}
\cmidrule{1-7}     \multicolumn{1}{c}{íon} & $\lambda_{obs}$ (\text{\AA}) & $\Delta v$ (km s$^{-1}$) & IP (eV) & FWHM & Flux  & Corrected flux &   &   &   &  \\
\cmidrule{1-7}       $[$S III$]$ & 9069.1  & 16.5   & 23.3  & 278$\pm$9    & 43.06$\pm$0.57  & 93.07$\pm$1.23  \\
Fe II                                             & 9181.7  & 44.9   & 7.9   & 276$\pm$56   & 5.86$\pm$1.19   & 12.46$\pm$2.53  \\
H I                                               & 9229.5  & -66.5  & 0.0   & 321$\pm$39   & 8.66$\pm$1.05   & 18.30$\pm$2.22  \\
$[$S III$]$                                       & 9531.1  & 0.0    & 23.3  & 279$\pm$9    & 101.08$\pm$0.56 & 205.51$\pm$1.14 \\
$[$C I$]$                                         & 9850.1  & -4.9   & 0.0   & 217$\pm$43   & 5.10$\pm$1.01   & 10.00$\pm$1.98  \\
$[$S VIII$]$                                      & 9914.0  & 30.2   & 280.9 & 284$\pm$12   & 12.33$\pm$0.52  & 23.96$\pm$1.01  \\
H I                                               & 10050.4 & 30.8   & 0.0   & 284$\pm$22   & 7.83$\pm$0.60   & 15.07$\pm$1.14  \\
\end{tabular}%
   \label{tab:esoinfra_tabelao}%
 \end{table*}%

\begin{table*}
   \centering
   \caption{Emission-line flux to the galaxy SDSS J164126.90+432121.5 in the NIR. FWHM is in units of km s$^{-1}$ and the flux in $10^{-16}$  erg s$^{-1}$ cm$^{-2}$}
     \begin{tabular}{lcccccccccc}
\cmidrule{1-7}     \multicolumn{1}{c}{íon} & $\lambda_{obs}$ (\text{\AA}) & $\Delta v$ (km s$^{-1}$) & IP (eV) & FWHM & Flux  & Corrected flux &   &   &   &  \\
\cmidrule{1-7}    $[$FeXI$]$ & 7884.6  & -273.5 & 262.1 & 770$\pm$165  & 2.85$\pm$0.61  & 4.23$\pm$0.85  \\
$[$OI$]$                                      & 8450.7  & 154.1  & 0.0   & 503$\pm$72   & 1.96$\pm$0.28  & 2.77$\pm$0.39  \\
$[$S III$]$                                   & 9072.0  & 112.4  & 23.3  & 804$\pm$79   & 6.23$\pm$0.61  & 8.44$\pm$0.82  \\
$[$S III$]$                                   & 9534.1  & 94.4   & 23.3  & 805$\pm$31   & 16.41$\pm$0.64 & 21.70$\pm$0.85 \\
$[$S VIII$]$*                                 & 9908.5  & -136.1 & 280.9 & 968$\pm$394  & 1.77$\pm$0.72  & 2.30$\pm$0.93  \\
$[$S VIII$]$                                  & 9917.0  & 121.0  & 280.9 & 149$\pm$66   & 0.25$\pm$0.11  & 0.32$\pm$0.14  \\
H I                                           & 10052.3 & 87.5   & 0.0   & 598$\pm$357  & 1.39$\pm$0.83  & 1.80$\pm$1.07  \\
\end{tabular}%
   \label{tab:j164nir_tabelao}%
 \end{table*}%

\begin{table*}
   \centering
   \caption{Emission-line flux to the galaxy III ZW 77 in the NIR. FWHM is in units of km s$^{-1}$ and the flux in $10^{-15}$  erg s$^{-1}$ cm$^{-2}$}
          \begin{tabular}{lcccccccccc}
\cmidrule{1-7}     \multicolumn{1}{c}{íon} & $\lambda_{obs}$ (\text{\AA}) & $\Delta v$ (km s$^{-1}$) & IP (eV) & FWHM & Flux & Corrected flux &   &   &   &  \\
\cmidrule{1-7}     O I & 8448.9  & 90.2    & 0.0   & 920$\pm$144  & 2.94$\pm$0.46  & 5.26$\pm$0.82  \\
$[$S III$]$                             & 9069.0  & 13.2    & 23.3  & 415$\pm$33   & 3.85$\pm$0.31  & 6.42$\pm$0.51  \\
$[$S III$]$*                            & 9074.5  & 195.0   & 23.3  & 1524$\pm$861 & 2.00$\pm$1.13  & 3.34$\pm$1.88  \\
H I*                                    & 9217.3  & -462.7  & 0.0   & 1691$\pm$393 & 5.47$\pm$1.27  & 9.00$\pm$2.10  \\
H I                                     & 9233.0  & 47.2    & 0.0   & 412$\pm$751  & 0.17$\pm$0.31  & 0.28$\pm$0.51  \\
$[$S III$]$                             & 9531.0  & -3.1    & 23.3  & 416$\pm$16   & 8.31$\pm$0.32  & 13.33$\pm$0.52 \\
$[$S III$]$*                            & 9536.5  & 169.9   & 23.3  & 1526$\pm$243 & 7.48$\pm$1.19  & 11.99$\pm$1.90 \\
\end{tabular}%
   \label{tab:iiizw77infra_tabelao}%
 \end{table*}

\begin{table*}
   \centering
    \caption{Emission-line flux to the galaxy MRK 1388 in the NIR. FWHM is in units of km s$^{-1}$ and the flux in $10^{-15}$  erg s$^{-1}$ cm$^{-2}$}
     \begin{tabular}{lcccccccccc}
\cmidrule{1-7}     \multicolumn{1}{c}{íon} & $\lambda_{obs}$ (\text{\AA}) & $\Delta v$ (km s$^{-1}$) & IP (eV) & FWHM & Flux & Corrected flux &   &   &   &  \\
\cmidrule{1-7}     $[$S III$]$* & 9069.9  & 43.0   & 23.3  & 986$\pm$202 & 11.74$\pm$2.41 & 13.81$\pm$2.83 \\
$[$S III$]$                                      & 9072.2  & 119.0  & 23.3  & 317$\pm$17  & 14.45$\pm$0.78 & 16.99$\pm$0.92 \\
$[$S III$]$*                                     & 9531.9  & 25.2   & 23.3  & 987$\pm$72  & 33.04$\pm$2.41 & 38.37$\pm$2.80 \\
$[$S III$]$                                      & 9534.2  & 97.5   & 23.3  & 318$\pm$10  & 31.14$\pm$0.78 & 36.16$\pm$0.91 \\
$[$S VIII$]$                                     & 9914.9  & 57.5   & 280.9 & 289$\pm$51  & 3.89$\pm$0.69  & 4.48$\pm$0.79  \\
H I                                              & 10052.1 & 81.5   & 0     & 356$\pm$55  & 3.24$\pm$0.50  & 3.72$\pm$0.57  \\
He II                                            & 10126.6 & 6.2    & 0     & 403$\pm$40  & 5.23$\pm$0.52  & 5.99$\pm$0.60  \\
\end{tabular}%
   \label{tab:mrkinfra_tabelao}%
 \end{table*}%

 \begin{table*}
   \centering
   \caption{Emission-line flux to the galaxy 2MASX J113111.05+162739 in the NIR. FWHM is in units of km s$^{-1}$ and the flux in $10^{-16}$  erg s$^{-1}$ cm$^{-2}$}
     \begin{tabular}{lccccccc}
\cmidrule{1-6}     íon & $\lambda_{obs}$ (\text{\AA})& $\Delta v$ (km s$^{-1}$) & IP & FWHM & Flux &   &  \\
\cmidrule{1-6}
$[$Fe XI$]$** & 7884.2  & -288.7 & 235.0 & 172$\pm$14   & 5.68$\pm$0.45   \\
$[$Fe XI$]$                                  & 7892.2  & 15.2   & 235.0 & 382$\pm$58   & 6.63$\pm$1.01   \\
$[$Fe XI$]$**                                & 7901.4  & 364.7  & 235.0 & 181$\pm$14   & 6.01$\pm$0.48   \\
$[$S III$]$                                  & 9069.9  & 43.0   & 23.3  & 569$\pm$118  & 10.81$\pm$2.25  \\
$[$S III$]$                                  & 9531.9  & 25.2   & 23.3  & 570$\pm$47   & 29.01$\pm$2.37  \\
$[$S VIII$]$                                 & 9913.1  & 3.0    & 280.9 & 535$\pm$201  & 5.82$\pm$2.19   \\
H I                                          & 10049.5 & 3.9    & 0.0   & 455$\pm$224  & 3.62$\pm$1.78   \\
\end{tabular}%
   \label{tab:j113tab}%
 \end{table*}%

\begin{table*}
   \centering
   \caption{Emission-line flux to the galaxy NGC 424 in the NIR. FWHM is in units of km s$^{-1}$ and the flux in $10^{-14}$  erg s$^{-1}$ cm$^{-2}$}
     \begin{tabular}{lcccccccccc}
\cmidrule{1-7}     íon & $\lambda_{obs}$ (\text{\AA}) & $\Delta v$ (km s$^{-1}$) & IP (eV) & FWHM & Flux & Corrected flux &   &   &   &  \\
\cmidrule{1-7}   $[$S III$]$ & 9068.6  & 0.0    & 23.3  & 729$\pm$20   & 4.36$\pm$0.12  & 5.95$\pm$0.16  \\
$[$S III$]$                                    & 9530.6  & -110.1 & 23.3  & 693$\pm$9    & 10.89$\pm$0.12 & 14.51$\pm$0.16 \\
$[$S VIII$]$                                   & 9911.6  & -42.3  & 280.9 & 742$\pm$109  & 1.22$\pm$0.18  & 1.60$\pm$0.24  \\
H I                                            & 10055.4 & 179.9  & 0     & 663$\pm$121  & 0.66$\pm$0.12  & 0.86$\pm$0.16  \\
H I*                                           & 10027.9 & -640.4 & 0     & 2280$\pm$518 & 1.76$\pm$0.40  & 2.29$\pm$0.52  \\
He II                                          & 10123.9 & -73.7  & 0     & 776$\pm$99   & 1.26$\pm$0.16  & 1.63$\pm$0.21  \\
$[$S II$]$                                     & 10286.2 & -15.4  & 10.4  & 541$\pm$61   & 0.62$\pm$0.07  & 0.80$\pm$0.09  \\
\end{tabular}%
   \label{tab:ngcinfra_tabelao}%
 \end{table*}%

%%%%%%%%%%%%%%%%%%%%%%%%%%%%%%%%%%%%%%%%%%%%%%%%%%

% Don't change these lines
\bsp	% typesetting comment
\label{lastpage}
\end{document}